\documentclass[12pt,preprint]{aastex}
\usepackage{graphicx}

\def\mb{\ifmmode {{\rm B_{435}}}\else
                ${\rm B_{435}}$\fi}
\def\mv{\ifmmode {{\rm V_{606}}}\else
                ${\rm V_{606}}$\fi}
\def\mi{\ifmmode {{\rm i_{775}}}\else
                ${\rm i_{775}}$\fi}
\def\mz{\ifmmode {{\rm z_{850}}}\else
                ${\rm z_{850}}$\fi}
                
\def\emnum{124\ }
\def\numltm0175{53\ } % number of emission line objects with M0<-17.5
\begin{document}
\title{Morphology and  evolution of emission line galaxies in the Hubble
Ultra Deep Field}
\author{N. Pirzkal\altaffilmark{1},C. Xu\altaffilmark{1},I. Ferreras\altaffilmark{2},S. Malhotra\altaffilmark{1},B. Mobasher \altaffilmark{1},,J. Rhoads\altaffilmark{1},A. Pasquali\altaffilmark{3},N. Panagia\altaffilmark{1},A. M. Koekemoer\altaffilmark{1},H. C. Ferguson\altaffilmark{1}, C. Gronwall\altaffilmark{4}
}
\altaffiltext{1}{Space Telescope Science Institute, 3700 San Martin Drive, Baltimore, MD21218, USA}
\altaffiltext{2}{Department of Physics and Astronomy, University College London, Gower Street, London WC1E 6BT}
\altaffiltext{3}{Institute of Astronomy, ETH H\"{o}nggerberg, 8093 Zurich, Switzerland}
\altaffiltext{4}{Department of Astronomy \& Astrophysics, Pennsylvania State University, 525 Davey Laboratory, University Park, PA 16802}


\begin{abstract}
We investigate the properties and evolution of a sample of galaxies
selected to have prominent emission lines in low-resolution grism
spectra of the Hubble Ultra Deep Field (HUDF).  These objects,
eGRAPES, are late type blue galaxies, characterized by small proper sizes
($R_{50} \leqslant 2 kpc$) in the 4350\AA\ rest-frame, low
masses ($5 \times 10^{9} M_\odot$), and a wide range of luminosities
and surface brightnesses. The masses, sizes and volume densities of
these objects appear to change very little up to a redshift of $z=1.5$.
On the other hand, their surface brightness decreases significantly from $z=1.5$ to
$z=0$ while their mass-to-light ratio increases two-folds. This could be a sign that most of low redshift eGRAPES have an older stellar population than high redshift eGRAPES and hence that most eGRAPES formed at higher redshifts. The average volume density of eGRAPES is $(1.8 \pm 0.3) \times
10^{-3} \ {\rm h}^3_{70}\ {\rm Mpc}^{-3}$ between $0.3 < z \leqslant
1.5$. Many eGRAPES would formally have been classified as Luminous
Compact Blue Galaxies (LCBGs) if these had been selected based on
small physical size, blue intrinsic color, and high surface
brightness, while the remainder of the sample  discussed in this paper forms an extension of LCBGs towards fainter luminosities.
\end{abstract}

\section{Introduction}
A majority of nearby galaxies are disk galaxies, still undergoing a significant amount of stellar formation. In order to understand how these objects have evolved towards their current shapes, sizes, and masses,  it is important to find out if these objects, as a population, have changed significantly over the last few billions years. Several groups have recently attempted to study disk dominated galaxies over a wide range of redshifts. \cite{ravindranath2003} observed that disk galaxies, selected from their broad band morphologies did not show a significant sign of size evolution over the redshift range of $0.25 \leqslant z \leqslant 1.25$. More recently,  \cite{barden2005} investigated the size evolution of disk galaxies up to $z=1$ and showed that these luminous objects ($M_v \leqslant -20.0$) showed a strong evolution in the magnitude-size relation, with an increase in surface brightness of 1 mag per square arc-second in the V band rest-frame. They also found these objects to have lower mass-to-light ratio at $z=1$ than at the present day and interpreted this as a lack of evolution between stellar mass and effective disk sizes in these objects. Their finding contradicts the results from \cite{ferguson2004} who found that the size of high redshift ($z \approx 1$ to $z \approx 5$) galaxies appear to evolve as $H^{-1}(z)$, in agreement with hierarchical formation theory.
In this paper, we examine the physical properties of  a new sample of star-forming galaxies. These objects were selected purely spectroscopically and,  unlike previous studies, allow us to   examine the morphology, size, surface brightness, and mass evolution of star forming galaxies without having to first pre-assume anything about the physical attributes  of these objects.
Are higher redshift star forming galaxies smaller, less massive than present day ones? Are their mass-to-light ratio significantly different than present day galaxies? In this paper, we examine these specific issues. \\

\section{Observations}
The Hubble Ultra Deep Field  \citep[HUDF][]{beckwith2005} is currently, and likely to remain for several years, the deepest set of observations ever taken of the sky. The Advanced Camera for Surveys (ACS) images of this field are extremely deep, containing over 10,000-15,000 galaxies down to a limiting magnitude of $\mz=29.5$ (AB magnitude), have  a small pixel scale (0.03''/pixel), and a very stable PSF. While the off-axis location of ACS in the HST focal plane results in a significant amount of image distortion, it is now a well calibrated and correctable effect (down to less than 0.1 pixel in the HUDF images). These images thus provide an excellent opportunity to measure the size and morphology of faint galaxies ($M_B \lesssim -18.0$), extending previous studies to much lower luminosities. While the HUDF images do not provide spectroscopic redshifts of individual sources, the GRAPES program  \citep[GRism ACS Program for Extragalactic Science, PI: Malhotra, see description in ][]{pirzkal2004} yielded slitless spectroscopic observations of 1400 objects with $\mi \leqslant 27.0$, or about 10\% of the sources in the HUDF). We identified  \emnum objects (eGRAPES) with prominent emission lines \citep{xu2005}. Ten eGRAPES are Lyman-$\alpha$ sources at $4.1 \leqslant z \leqslant 5.8$  \citep{xu2005}, while the bulk of eGRAPES actually lay at redshifts of $0. \leqslant z \lesssim 1.5$.\\
Since the HUDF images reach 2.5 magnitudes deeper than the limiting magnitude of the GRAPES catalog, all of the objects in the GRAPES catalog have very high signal-to-noise images in the \mb, \mv, \mi, and \mz\ HUDF ACS observations (corresponding roughly to the B,V,i, and z bandpasses). We estimate the limiting surface brightness of the HUDF images to be 25.7, 26.5, 25.6, and 24.9 magnitude per square arc-second. In addition to this, near infrared observations of the HUDF have also been available \citep{thompson2005} and offer J and H band observations for a large number of these objects, albeit at a much lower resolution (0.09''/pixel).  This allows for accurate measurements of the magnitude, shape and size of these objects to be made in different rest-frame wavelengths.

\section{Sample and measurements}\label{meas}
For the purpose of studying variation in physical properties of eGRAPES as a function of redshift, it is useful to start by defining the following redshift bins, each initially containing an almost equal number of eGRAPES ($\approx $20--30): $0. \leqslant z \leqslant  0.3$, $0.3 \leqslant z \leqslant  0.55$, $0.55 \leqslant z \leqslant  0.85$, $0.85 \leqslant z \leqslant  1.5$. The average (median) redshift of objects in each of these bins is $z=0.20 (0.20), 0.41 (0.41), 0.73 (0.73), 1.15 (1.11)$ respectively. In these four bins, the {4350\AA } rest-frame corresponds to the observed wavelengths of 5220\AA, 6133\AA, 7526\AA, and 9353\AA, nearly equivalent to the \mb, \mv, \mi,{ \mz\ } ACS bandpasses. Throughout this paper, we computed {4350\AA } rest-frame parameters (magnitude, sizes, morphologies) by linearly interpolating between measurements of these quantities made separately in the two closest available ACS bands.\\
The initial number of eGRAPES is relatively small (124), but we must further impose a luminosity cut on our sample so that measurements at lower redshifts we make are not biased by less luminous, smaller galaxies whose counterparts would not have been detected at higher redshifts. Assuming a concordant cosmology with $\Omega_M = 0.3, \Omega_\Lambda = 0.7, {\rm and\ } h=0.7$, the GRAPES limiting magnitude of \mi=27 implies that only objects brighter than $M_\mb \approx -17.5$ would be observed at $z=1.2$. Based on simulations, we estimate that more than 90\% of emission lines with an line equivalent-width greater than 75\AA\ were detected \citep{xu2005}. Applying this luminosity cut further lowers the number of emission line objects down to \numltm0175, and the number of objects remaining in each of our redshift bin is 5, 9, 13, 27.

In this paper, we also carried out a parallel analysis of non-emission line galaxies with known photometric redshifts. The redshifts of these objects were determined by the GOODS project using the combined ACS (BViz) and NICMOS (JH) data from the HUDF and deep K\_s band (ISAAC)
data. All these data were degraded to the ISAAC seeing (0.4 arcsec) and combined to estimate photometric redshifts by fitting their SEDs to those of rest-frame templates. The photometric redshifts of the non-emission line objects presented in this paper, measured using luminosity function priors, have an accuracy of $(z_{spec}-z_{phot})/(1+z_{spec}) \approx 0.08$ \citep{mobasher2005}. A detailed analysis of these non-emission line galaxies is beyond the scope of this paper and we only use these objects to serve as a comparison sample against which we can compare our eGRAPES sample.

There are several well established methods to measure the size and shape of galaxies and we studied  the morphology of eGRAPESs using three separate methods. First, we measured the apparent size (half-light radius, $R_{50}$) and magnitude each eGRAPES in the \mb, \mv, \mi, and {\mz\ } bands using the SExtractor program \citep{Sextractor} (in dual image mode using a combination of the  {\mi\ } and {\mz\ } band images as the detection image). \\
Second, we also used the program GALFIT \citep{peng2002} to fit a S\'ersic profile \citep{sersic1968} to each objects. This profile is of the form $\Sigma(r) = \Sigma_e e^{-k {[r / r_e}^{1/n} - 1]}$, where $r_e$ is the effective radius of the source, $\Sigma_e$ is the surface brightness at $r_e$, $n$ is the power-law index, and $k = k(n)$ is a normalization constant. For $n=4$, the S\'ersic profile reduces to a classic de Vaucouleurs profile while for $n=1$ it reduces to an exponential disk profile. We used segmentation maps from SExtractor  as input masks when running GALFIT and we started each fit  with an initial value of $n=4$. While many eGRAPES appear to be relatively small on the ACS images ($\approx 5$ HUDF pixels), this did not affect the SExtractor and GALFIT size estimates and both agreed reasonably well. \\
Third, we computed the  Concentration (C) and Asymmetry (A) values \citep{conselice2000} of eGRAPESs.
Measuring A consists of rotating and subtracting a galaxy image from itself and computing the sum of the absolute value of the residuals. Measuring C consists in computing the logarithm of the ratio of the radii enclosing 20\% and 80\% of the light in the object. The relations between A, C, and galaxy morphology have been extensively studied in the past  \citep{conselice2000, conselice2003} and have been well calibrated using a sample of nearby objects \citep{bershady2000}.  This last method proved to be the most difficult one to apply to eGRAPES which appear as small objects in the HUDF. 
\citet{conselice2000} estimated that a physical image resolution of 0.5 kpc was acceptable in order to produce reliable measurements of A, and they defined the parameter $\epsilon = {\theta_{0.5kpc} \over \theta_{res}}$. As $\epsilon$ falls significantly under unity, CAS parameters, and the asymmetry parameter A in particular, become less reliable.  \citet{lauger2004} showed that they were able to successfully measure CAS values of objects with  $\epsilon \geqslant  0.6$. Additionally, the reliability of CAS measurements of objects with different surface brightnesses was investigated by \cite{lauger2004}. The authors  concluded that a signal-to-noise ration per pixel ${S\over{N_{pix}}} \geqslant 1$ was required for unbiased asymmetry measurements  (A lower number of high signal-to-noise pixels, as diffuse extended regions of objects become invisible, cause A to be computed using a smaller number of pixels which then artificially increases the estimate of A. See Figure 1 in \citet{lauger2004}).

The HUDF ACS z band images have a resolution of 0.084'' (combining the pixel scale and the size of the ACS PSF), corresponding to  $\epsilon \ge 0.7$ for $ 0 \leqslant z \leqslant 6$, which, while lower than the value recommended by \citet{conselice2000}, is larger than the accepted value used by \citet{lauger2004}.  We nevertheless independently investigated the effect that small eGRAPES sizes might have on our CAS measurements of these objects. We started by selecting 166 large and bright objects and measured their CAS values after scaling down the size of the images by factors of 2, 3, 4, 5 and 6.  We found that A measurements remained robust, and C measurements remained mostly unaffected as long as objects had a half-light radius (R\_{50}) that is equal or greater than 5 HUDF pixels. We computed the $S\over{N_{pix}}$ for eGRAPESs  in the filter closest to the 4350{\AA } rest frame wavelength and verified that the high-signal to noise ACS imaging ensured that ${S\over{N_{pix}}} \geqslant 1$ for these objects. 

%\begin{figure}
%\includegraphics[width=7.0in]{fig1}
%\caption{The redshift distributions of the emission lines, V dropouts, and I dropouts objects discussed in this paper.\label{fig1}}
%\end{figure}

\section{Nature of eGRAPES}
\begin{figure}
\includegraphics[width=7.0in]{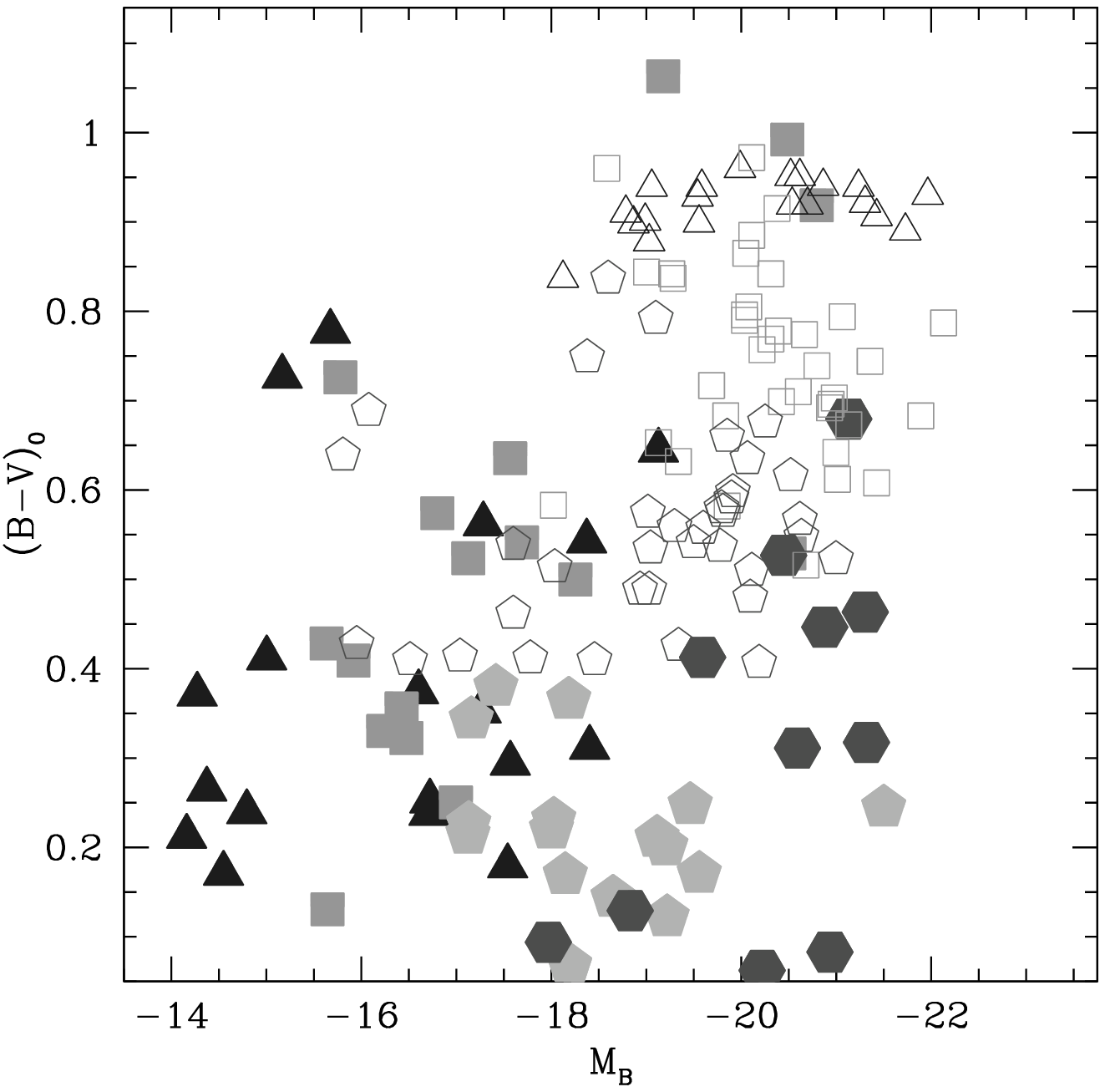}
\caption{Rest-frame (\mb\ - \mv) vs rest-frame $M_\mb$. The empty symbols are low redshift galaxies from \citet{bershady2000}, \citet{frei1996}, and \citet{kent1984}. E-S0, Sa-Sb, and Sc-Irr are shown using triangles, squares, and pentagons, respectively. The solid symbols represent eGRAPES objects. eGRAPES at $ 0 < z \leqslant 0.3 $, $ 0.3 \leqslant z \leqslant 0.55 $, $ 0.55 \leqslant z \leqslant 0.85 $, and $ 0.85 \leqslant z \leqslant 1.5 $ using triangles, squares, pentagons, and hexagons, respectively. Most eGRAPES have bluer rest-frame colors and lower luminosities than even the  local Sc-Irr objects shown in \cite{bershady2000}.\label{morph1}}
\end{figure}

\begin{figure}
\includegraphics[width=7.0in]{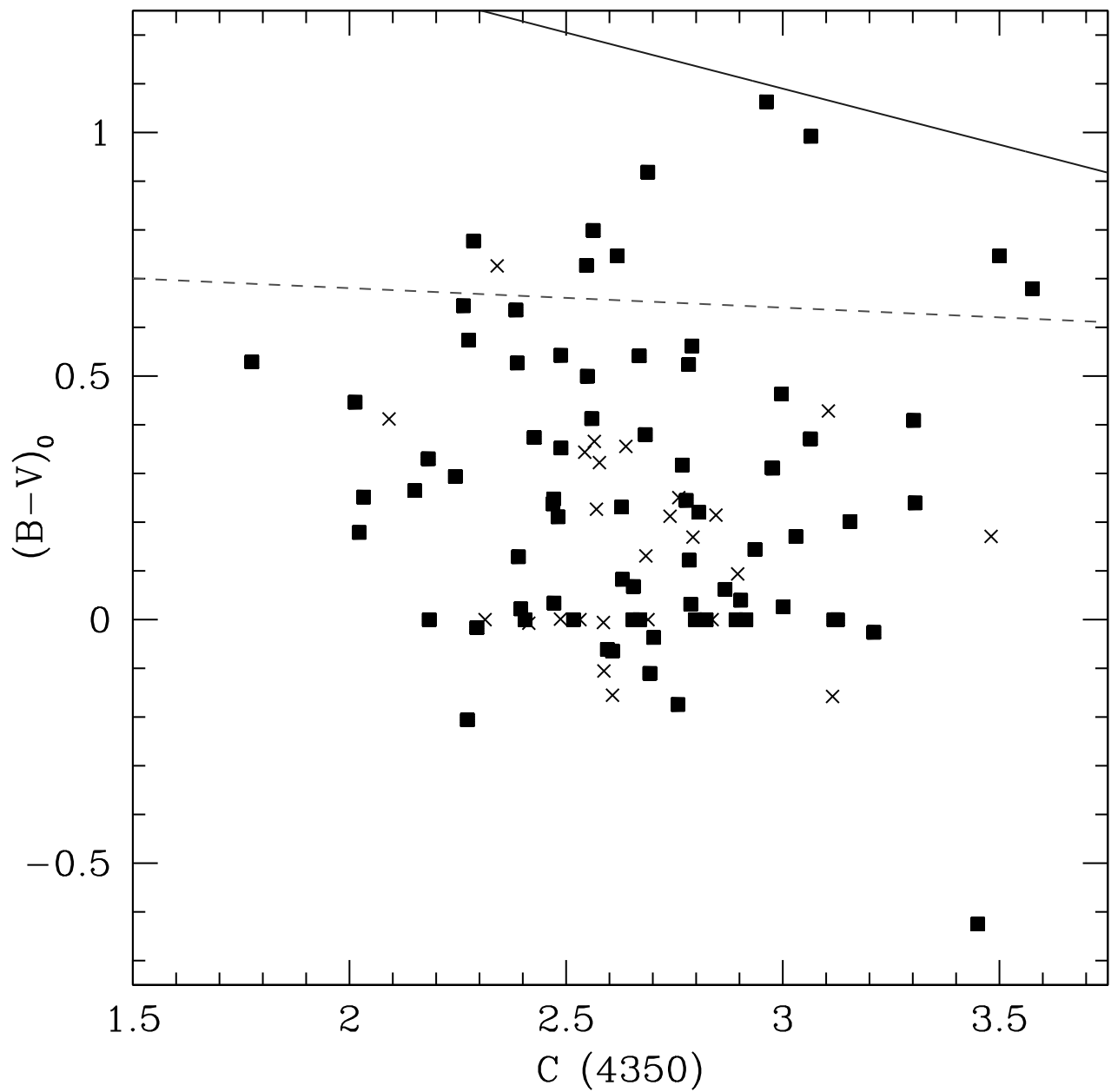}
\caption{Rest-frame (\mb\ - \mv) vs Concentration for eGRAPES. Late type objects are in the region under the dash curve. Early type objects are above the solid curve. Intermediate objects are between the two curves \citep{bershady2000}. The crosses are objects with marginal surface brightness and/or that are smaller than 5 HUDF pixels and for which CAS values might be suspect. 90\% of eGRAPES are late type galaxies.\label{fig3}}
\end{figure}

\begin{figure}
\includegraphics[width=7.0in]{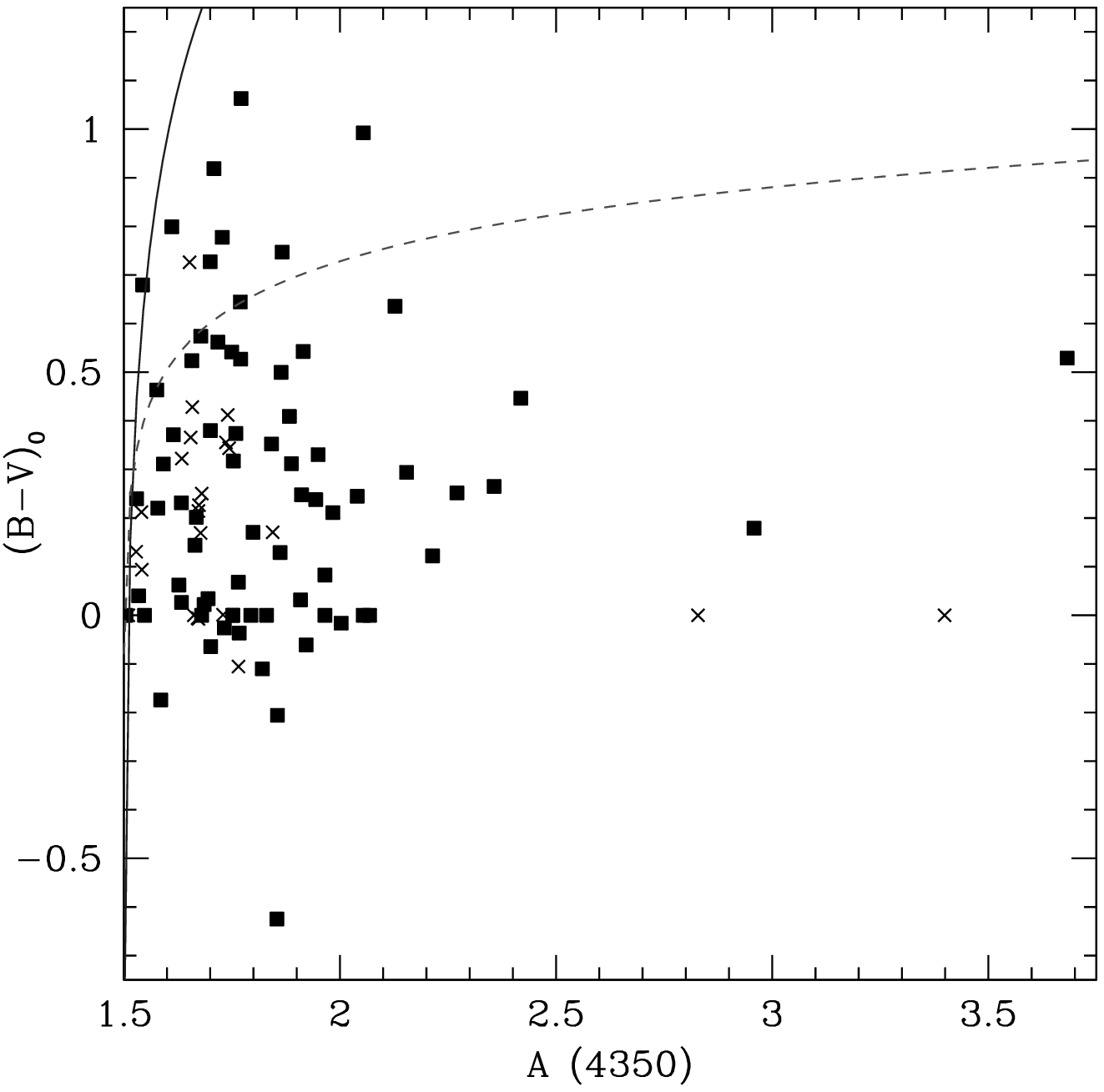}
\caption{Rest-frame (\mb\ - \mv) vs Asymmetry. Late type objects are in the region under the dash curve. Early type objects are above the solid curve. Intermediate objects are between the two curves \citep{bershady2000}. The crosses are objects with marginal surface brightness and/or that are smaller than 5 HUDF pixels and for which CAS values might be suspect. \label{fig4}}
\end{figure}

Since eGRAPESs were not selected based on their physical attributes, we did not a-priori attempt to select objects with late-type colors and morphologies.
One can however distinguish galaxies of different spectral types using a simple rest-frame color versus absolute magnitude plot as well as color versus CAS parameter plots.  \citet{bershady2000} showed that determining the spectral type of objects in this manner remains consistent across a broad range of redshifts and angular sizes. Figure \ref{morph1} shows the eGRAPES population  plotted in a rest-frame (\mb - \mv) versus rest-frame $M_{\mb}$\ magnitude diagram.  It is clear from Figure \ref{morph1} that eGRAPESs are extremely blue compared to local, present day galaxies, and that nearly all eGRAPES are even bluer than local Sc-Irregular galaxies. Figures \ref{fig3} and \ref{fig4}, confirm that the spectroscopically selected eGRAPESs preferentially occupy the late and intermediate parts of rest-frame (\mb - \mv)  versus C and A plots, respectively. In fact, 90\% of eGRAPES would be classified as late type if one were to rely on these plots. The high fraction of late-type galaxies amongst eGRAPES has also been confirmed by the work of  \citet{mobasher2005} who assigned irregular or starburst  spectral types to  68\% of eGRAPES based on photometric band fitting.

\subsubsection{Morphology as a function of redshift}\label{morph}
%We also investigated the redshift dependence, for $ z \leqslant 1.5$, of the asymmetry (A) and concentration (C) of eGRAPES in the \mb\ rest-frame. Figure \ref{fcasa} shows the redshift dependence of A. As discussed earlier, we restricted our sample of eGRAPES to objects large enough and with a surface brightness that is high enough to allow for unbiased CAS measurements. Figure \ref{fcasa} shows that low redshift eGRAPES are not significantly more asymmetric or concentrated than higher redshift ones. 

%As discussed in Section \ref{meas}, and while measuring CAS parameters for these objects is rendered difficult by their small apparent sizes and finite image resolution,  eGRAPES object should be sufficiently, if somewhat marginally,  resolved to produced meaningful A and C measurements and any large redshift trend in A or C values of these objects would have been detected. 

% In any case, and as discussed above, and as shown in Figures \ref{AR50} and \ref{CR50}, the failure to properly resolve these objects at increasing redshift would actually cause A and C values to strongly decrease as a function of redshift, which is not observed here.\\

\begin{figure}
\includegraphics[width=7.0in]{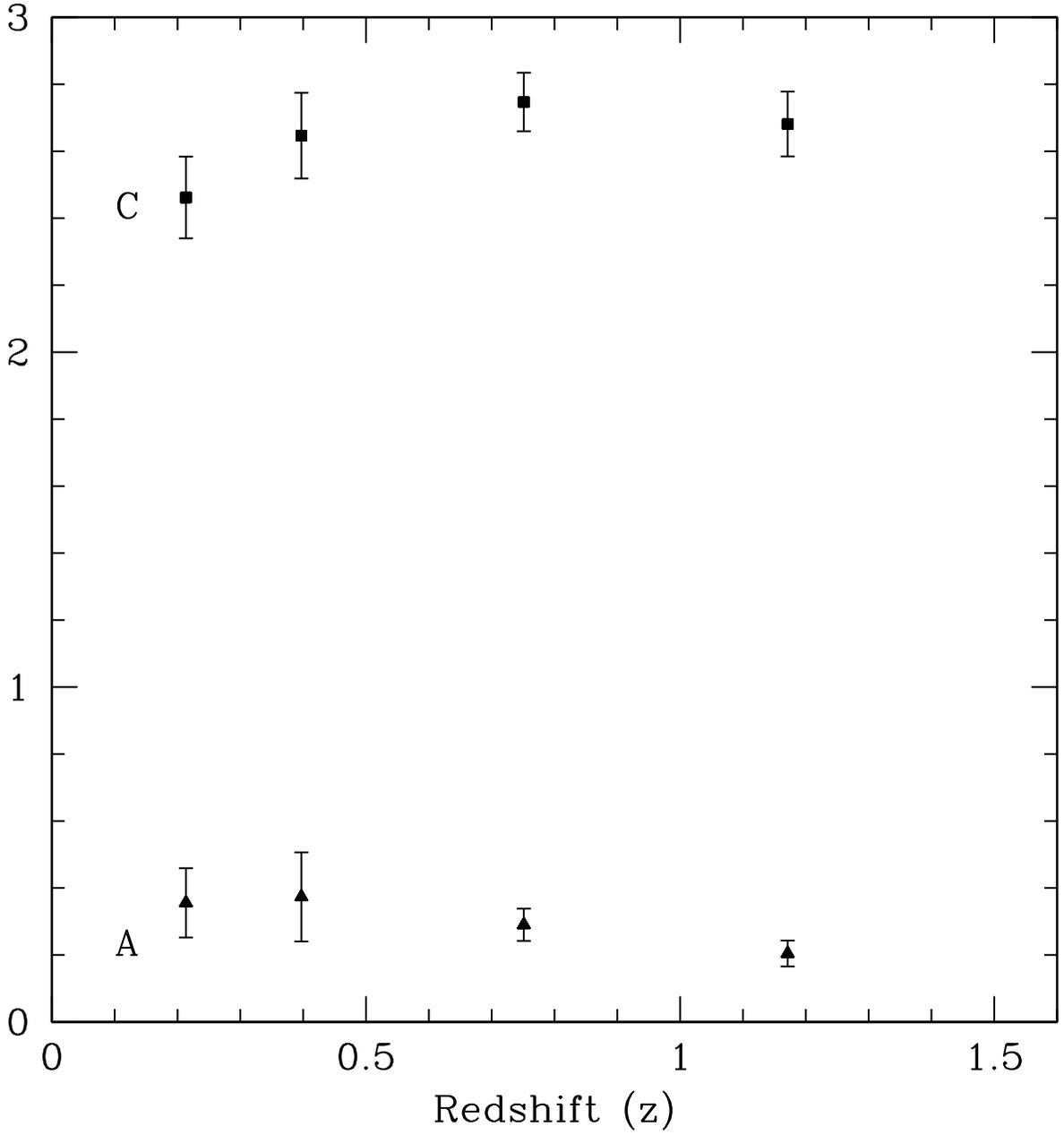}
\caption{4350\AA\ rest-frame asymmetry (A, bottom points) and concentration (C, top points) values for eGRAPES with $M_{\mb} \leqslant -17.5$. These are the values averaged in each of redshift bins, and the error bars are the standard deviation of the mean. Neither the asymmetry nor the concentration of eGRAPESs  vary significantly from  $z \approx 1.5$ to the present day.  \label{fcasa}}
\end{figure}

\begin{figure}
\includegraphics[width=7.0in]{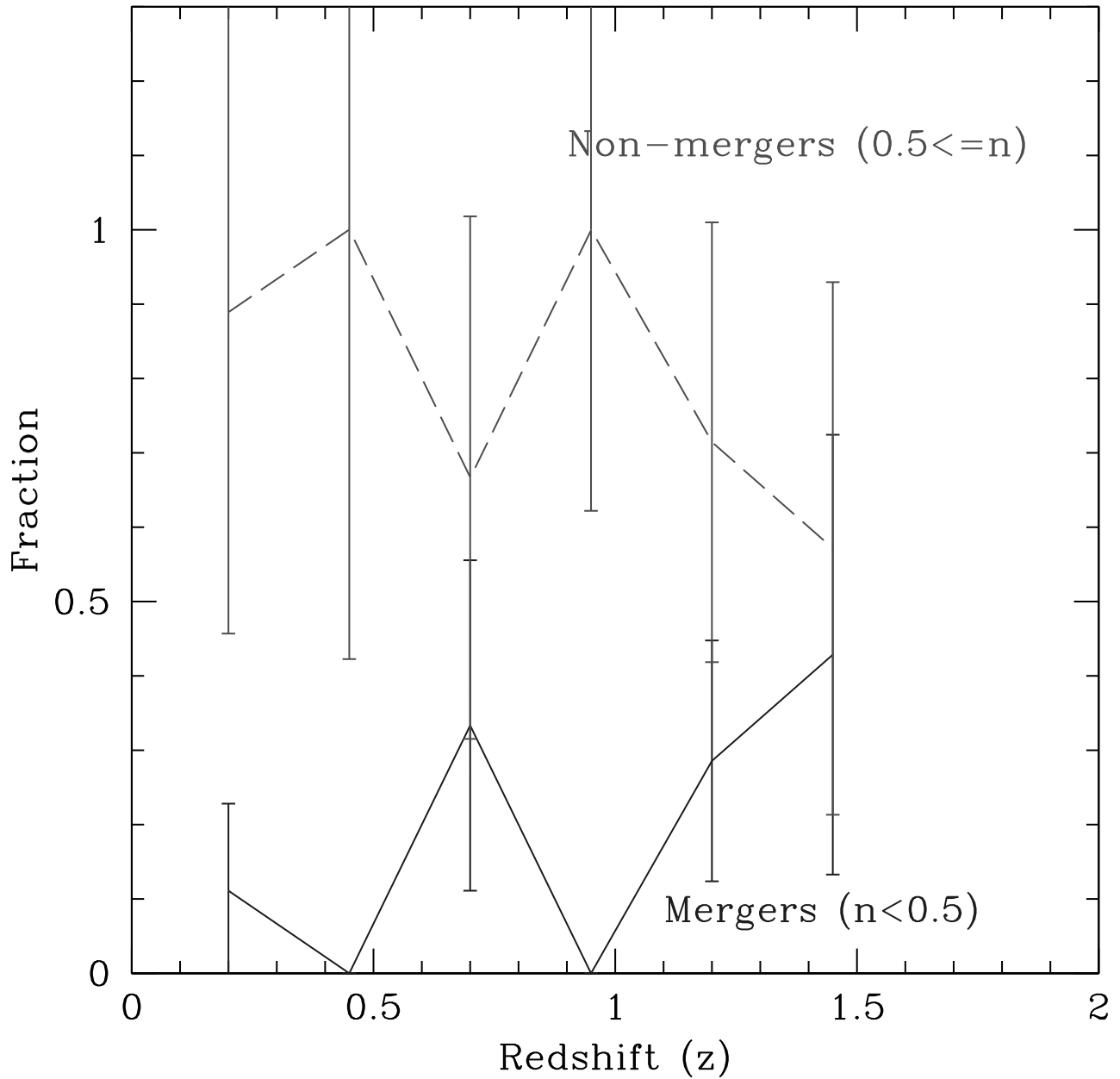}
\caption{Redshift evolution of the ratio of merger, and non-merger eGRAPES to the total number of eGRAPES, as a function of redshift.  Only objects with $M_{\mb} \leqslant -17.5$ are included. eGRAPES best fitted by a S\'ersic index smaller than 0.5 are classified as mergers. The large error bars ($1 \sigma$) reflect the small number of available sources. The overall fraction of mergers (i.e. significantly flatter profiles than exponential profiles) is small at all redshifts and does not appear to vary significantly as a function of redshift. \label{galfitratio}}
\end{figure}

We classified eGRAPESs in separate morphological groups based on our Galfit S\'ersic fit: mergers ($n<0.5$), and non-mergers ($n>0.5$). We assumed that most merger system have in fact light profiles which are significantly flatter than pure exponential profiles \citep{marleau1998}.
We found that 11 objects with $M_\mb \leqslant -17.5$ qualify as mergers (19\% of total) and furthermore plot the fraction of merger candidate as a function of redshift in Figure \ref{galfitratio}. No strong trend is visible in  Figure \ref{galfitratio} and the fraction of eGRAPES merger candidates remains small at all reshifts except near $z=1.5$ where the merger fraction increases marginally. The number of available objects is small however and the errors are dominated by small number statistics and a larger population of eGRAPES is required to confirm this observation.\\
We also computed the asymmetry (A) and concentration (C) of eGRAPESs in the \mb\ rest-frame. As discussed earlier, we restricted the CAS analysis to eGRAPESs that are large enough and that have a surface brightness that is high enough to allow for unbiased CAS measurements. Figure \ref{fcasa} shows that neither the asymmetry nor the concentration of these objects appear to evolve significantly as a function of redshift. \\
One can only conclude at this point that there does not appear to be a significant change in the average morphology of eGRAPESs, as a population, from the redshift of $z \approx1.5$ to the present day. However, the number of eGRAPESs for which these measurements could be performed in an unbiased way is small and a much larger sample would be required to definitively state that these objects  did not change much, as a population.

\section{Size Evolution}
\begin{figure}
\includegraphics[width=7.0in]{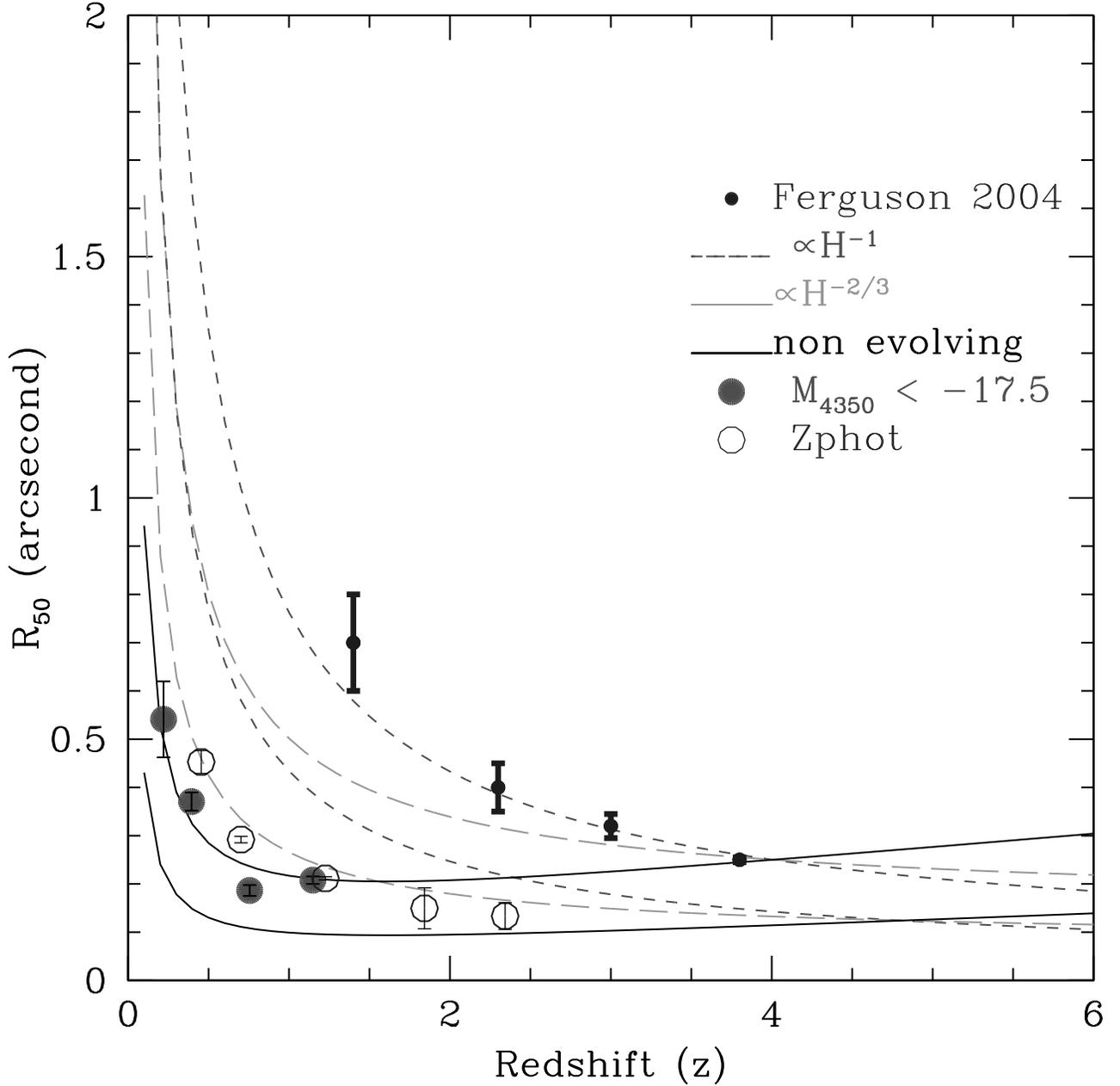}
\caption{Measured, apparent sizes (in '') of eGRAPES in the 4350\AA\ rest-frame as a function of redshift. eGRAPES  are plotted in large filled circles. The sizes of \citet{ferguson2004} disk galaxies are plotted in small solid circles. The error bars we show for the eGRAPES points are the 95\% confidence limit of our measurements. The errors bars from the \citet{ferguson2004} data are their original error bars which were determined using simulations. The solid lines show the expected size distribution in the case of non-evolution. The long dash and short dash curves are as in \citet{ferguson2004} and show the expected evolution if sizes scaled as the halo masses ($R \propto H^{-1}(z)$  for disks with fixed circular velocity (red)  or  and $R \propto H^{-{2\over3}}(z)$ for fixed mass (green)). The various evolution models are shown using two different normalization at $z=4$.\label{fig2}}
\end{figure}

\begin{figure}
\includegraphics[width=7.0in]{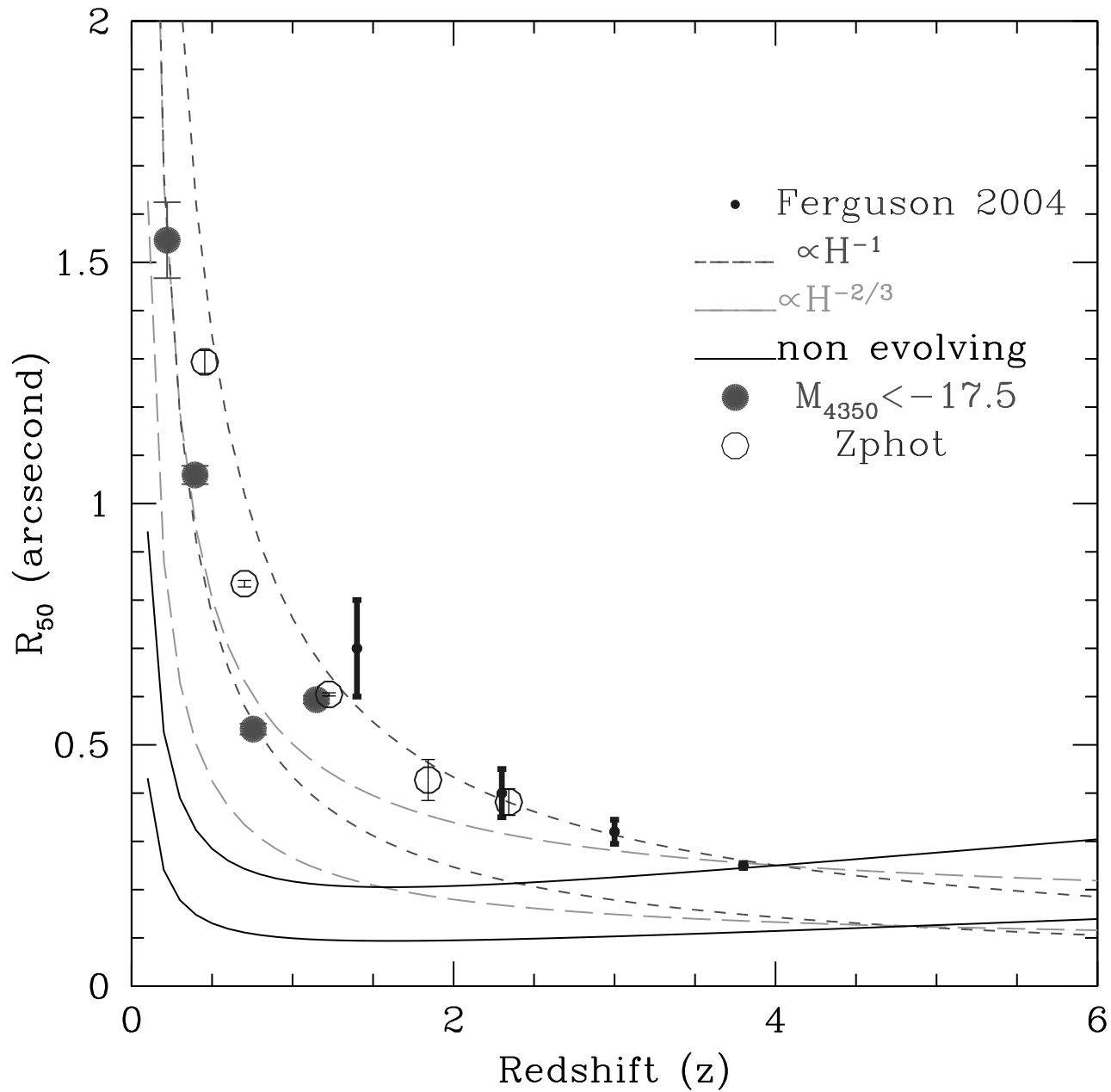}
\caption{Estimated, apparent sizes (in '') of eGRAPES in the 1500\AA\ rest-frame as a function of redshift. eGRAPES  are plotted in large filled circles. Symbols and lines are as in Figure \ref{fig2}..\label{fig2b}}
\end{figure}

We also examine the physical sizes of eGRAPES (half-light radius, $R_{50}$) as a function of redshift to shed some light on the evolution of this group of star forming galaxies over the last 8 billions years. We computed the apparent {\mb\ } rest-frame sizes of eGRAPESs (in arc-seconds) by interpolating the  half-light radii measured individually using SExtractor in the  \mb, \mv, \mi, {\mz\ } bands.  Figure  \ref{fig2} shows the eGRAPES sizes as a function of redshift for objects with $M_\mb \leqslant -17.5$, as well as the sizes of photometric redshift galaxies in the field. \\
\citet{ferguson2004} showed a sample of what the authors assumed to be disk supported galaxies to be evolving roughly as $\propto H^{-1}$, where $H$ is the Hubble parameter, in agreement with what is expected if the \citet{fall1980} model of disk formation with fixed disk circular velocity within dark matter halos is correct. At first glance, it appears that eGRAPES sizes evolve very little as a function of redshift, and certainly less than disk dominated galaxies from previous studies such as  \citep{ferguson2004}.  Our control sample of non-emission line, photometric redshift galaxies appear to have sizes that closely follow a $H^{-1}(z)$ evolution model but have sizes that are overall still much smaller than the data from \citep{ferguson2004}. \\
Several effects have to be taken into account however before any conclusion can be made. First,  eGRAPESs were selected spectroscopically based on the presence of  emission lines in their slitless spectra \citep{pirzkal2004, xu2005}. The resolution in such spectra is directly affected  by the physical size of the underlying object \citep{pasquali2005} which could lead to some biases against the detection of larger objects. As objects get larger, the effective resolution of the grism observations is reduced and the sharpness of any emission line in the spectra is lowered. We estimated this effect using a Monte-Carlo simulation containing 100,000 simulated spectra with a wide range of line equivalent widths, objects sizes and  brightnesses. The underlying size distribution of these simulated objects was taken directly from \citet{ferguson2004} at redshifts of $z=1.4$ and $z=2.3$ (with average sizes of  0.6'' and 0.3'', respectively). Our simulations showed that most of the emission lines in these simulated spectra, even when they corresponded to objects that were significantly larger than eGRAPESs, were properly detected.  The resulting average sizes for the simulated eGRAPESs were 0.5'' and 0.3'' in the $z=1.4$ and $z=2.3$ bins respectively, hence demonstrating that our eGRAPES sample is not strongly biased towards small sources.\\
Another thing to keep in mind is that \citet{ferguson2004} selected objects using a rest-frame UV luminosity cut corresponding to $M_{AB}(1700\AA)=-21.0$ for a $z\approx4$ LBG \citep{steidel1999}, while the small area covered by the HUDF forces us to select objects with $M_\mb$=-17.5. We did estimate the impact that our luminosity cut has on the average sizes by computing new mean sizes using a series of luminosity cuts: Objects with $0.7 L* \leqslant L \leqslant L*$, where L* is the luminosity of galaxies with absolute magnitudes M*, were selected while M* was varied over a large range of possible values  ranging from -23.0 to -16.0.  As increasingly brighter objects were selected using increasingly high luminosity cuts, the average sizes of eGRAPES in each redshift bin increased only marginally (while the number of eGRAPES left in each redshift bin decreased significantly). At $z=1.15$ (which is within the range of redshifts probed by \citet{ferguson2004}), the average size of objects increased  from  0.2'' at M=-16 to 0.27'' at M=-20.5. The impact of our lower luminosity cut is therefore not likely to be the cause of the large observed size differences shown in Figure \ref{fig2}.\\
The last thing to realize  when comparing eGRAPES sizes to the ones from \citet{ferguson2004} is that our measurements were made in the \mb\ rest-frame while \citet{ferguson2004} used the 1500\AA\ rest-frame. While this has the advantage to allow us to measure the radial extend of these objects based on their global stellar mass distribution rather than based on the  instantaneous, unobscured, star formation as measured  in the UV \citep{trujillo2004}, it  makes it difficult to compare our results  to the \citet{ferguson2004} results. In order to properly compare the two, we must estimate the difference in sizes of these objects when observed in the 1500\AA\ and 4350\AA\ rest-frames.  We estimated the magnitude of this effect using the sample of galaxies with known photometric redshift from \citet{mobasher2005} that are at $z \approx 1.9$ and that were successfully detected in the ACS \mb\ band as well as the NICMOS J band observations of the HUDF. Measuring the sizes of these objects  in both the \mb\ band and the J band, allowed us to directly measure their sizes in both the 1500\AA\ and 4350\AA\ rest-frames, respectively. Sizes were measured using the same method described above in both the ACS \mb\ band image and the NICMOS J band image. We first matched the resolution of the ACS \mb\ band image to the resolution of the NICMOS image using stars in the HUDF \citep{pirzkal2005}. We then computed the average ratio of sizes between the degraded ACS B band image (i.e. 1500\AA\ rest-frame) and the NICMOS J band image (i.e. 4350\AA\ rest-frame). We found that the average size ratio is $\approx 0.4$, a value that is somewhat larger  than what was observed by \citet{barden2005}  who estimated this bandpass effect to be only 20\% at $z\leqslant 1$, using GEMS galaxies. Figure \ref{fig2b} shows the estimated eGRAPES 4350\AA\ rest-frame sizes compared to the \citet{ferguson2004} data.  By applying this correction to the photometric redshift galaxy sample  the size distribution of these objects  agree well with the result  from \citet{ferguson2004}.  On the other hand, eGRAPES  remain smaller at all redshifts, even after correction for wavebands.

%While most of the GRAPES objects are intrinsically less bright than the lower luminosity cut that was %imposed on the \citet{fergsuon2004} sample, there  are eleven objects in the redshift bin %$0.85\leqslant z \leqslant 1.5$ with an absolute UV rest-frame luminosity $0.7 L_* \leqslant L \leqslant %5 L_*$. These objects, which have intrinsic luminosities as high as in the \citet{ferguson2004} sample, %have an average size of $0.23'' \pm 0.03$.  

\begin{figure}
\includegraphics[width=7.0in]{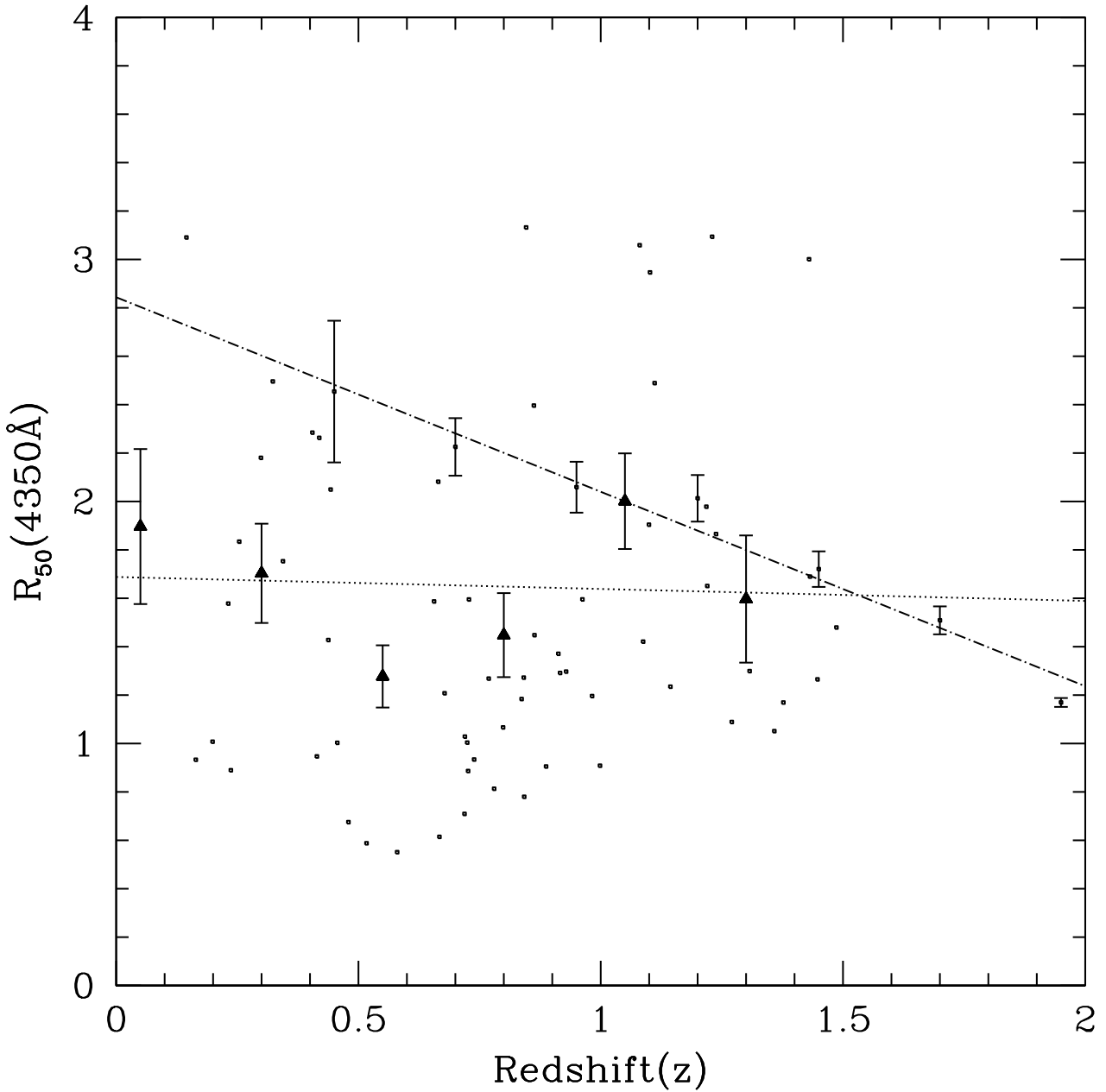}
\caption{Physical size (in kpc) of the $M_{\mb} \leqslant -17.5$ eGRAPES  in the \mb\ rest-frame. The sizes of non emission line, photometric redshift galaxies are also shown (dot-dash line) for comparison. Compared to non emission line galaxies, eGRAPES are observed to have a much more heterogeneous distribution of sizes and show little evidence of a strong  redshift-size relation. The dot-dashed lines show the least square fits to the sizes of eGRAPES and photometric redshift galaxies. Errors bars are the standard deviation in each of the chosen redshift bins. \label{R50zm17}}
\end{figure}

We compare the size evolution of eGRAPES to that of non-emission line photometric redshift objects by plotting the proper physical size of these objects (in kpc) as a function of redshift, as shown in  Figure  \ref{R50zm17}.
If sizes increases as $\propto H^{-1}$, one would expect that the proper sizes of these objects to monotonically increase by a factor of 2 between $z=1.15$ and $z=0.20$. Such a trend is clearly not present in Figure  \ref{R50zm17}  and no significant evolution  (i.e. $> 1 \sigma$) is observed. The least square fit of the sizes of eGRAPES is $R_{50}(z) = -0.05\times z + 1.69 {\rm\ kpc}$.  The proper sizes of the non star forming galaxies with known photometric redshift are also plotted (dot-dashed line) and show a strong redshift dependence, as expected \citep{ferguson2004,barden2005}. We estimated the effect of cosmological dimming on our size measurements using simulations and one would expect sizes to vary by a factor of two between the reshifts of z=0.3 and z=1.5, While it is difficult to apply this effect as a correction to our measurements, we can conclude that eGRAPES at low redshifts are certainly no larger than those at higher redshifts and might even have been larger in the past, but this claim would have to be verified with further observations and an larger number of eGRAPES. One is left with the overall picture that eGRAPESs are a population of objects heterogeneous in nature but whose redshift evolution, at least as far as proper sizes are concerned, may not be  coupled to the underlying dark matter halo distributions, unlike non-eGRAPES objects. The small field area of the HUDF and the small number of sources does however make this result sensitive to cosmic variation and small number statistics. Further HST/ACS grism observations, over a wider area, such as from the PEARS project (Probing Evolution And Reionization Spectroscopically, PI: Malhotra) should allow us to confirm this result in the future.

%At higher redshifts ($z \geqslant 3$)  the 1500{\AA\ } rest-frame sizes of these objects can be computed and directly compared  to the sizes shown in \citet{ferguson2004} and no size conversion factor needs to be applied. As shown in Figure \ref{fig2}, the GRAPES objects sizes are consistent with previous observations, simply confirming the known fact that high redshift LBGs have small physical sizes. The eGRAPES actually contains two types of objects at $4 \leqslant z \leqslant 5$, one being the {\mv\ } dropouts and the second one being the small group of objects with Lyman-$\alpha$ emission. A more detailed comparison of these objects will be the subject of a subsequent paper.

\begin{figure}
\includegraphics[width=7.0in]{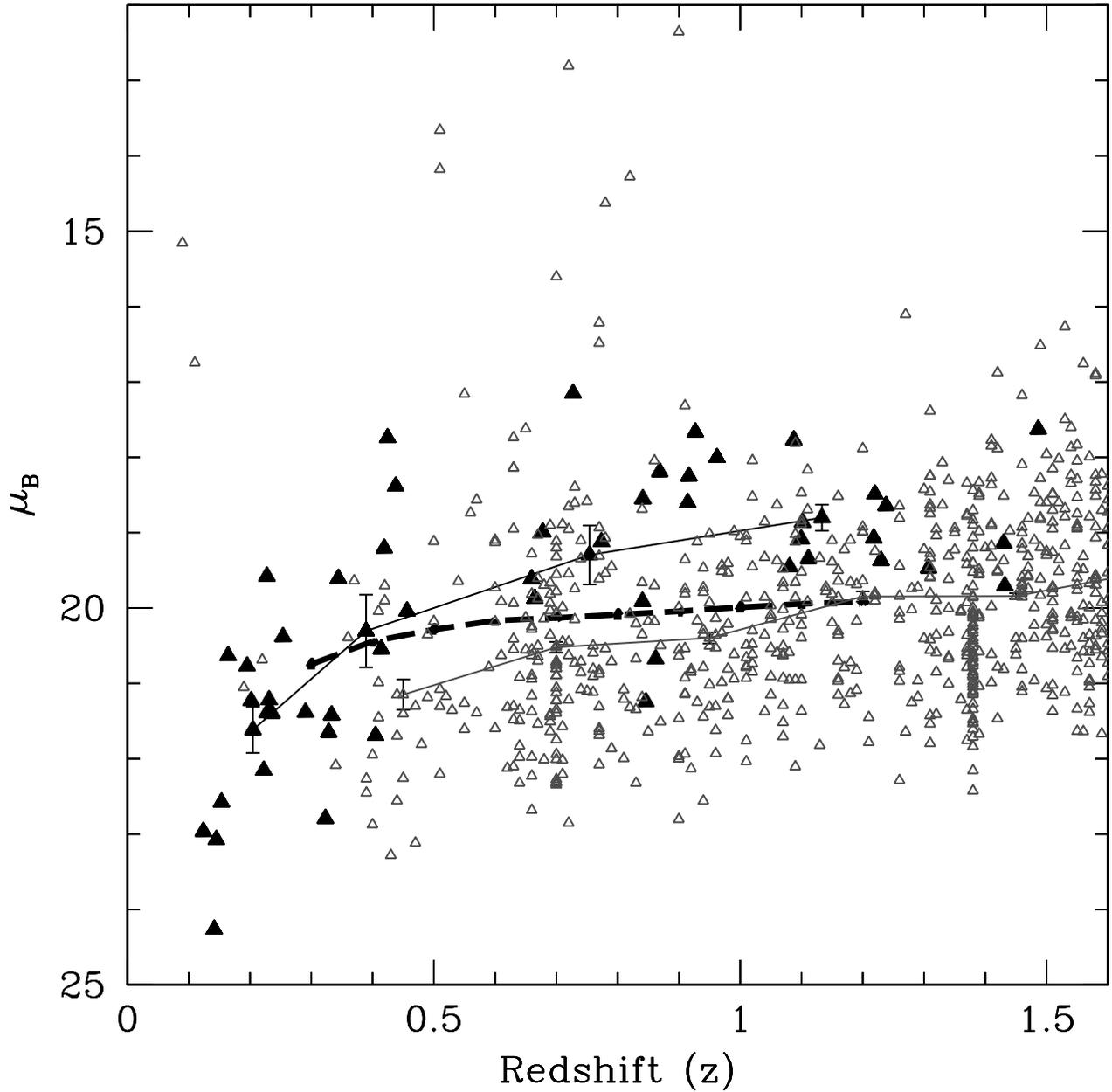}
\caption{Surface brightness of eGRAPES  (solid triangles) and non emission line photometric redshift galaxies (open triangles) as a function of redshift. Average values at redshifts of  $z=0.20, 0.41, 0.73, 1.15$ are also shown. The mean surface brightness is observed to decrease by $\approx 2.0$ magnitudes per arc-second$^2$ from z=1.15 to z=0.2. The dash line shows the simulated effect of cosmological dimming on effective surface brightness measurements. The observed changes in surface brightness as a function of redshift is stronger than the simulations.\label{fig6}}
\end{figure}

Related to the size evolution of eGRAPES, is the question of surface brightness evolution of these objects. Following \citet{barden2005}, we can compute the effective surface brightness of these objects in the rest-frame B band using:
\begin{equation}
\mu_B = M_b + 5 \log{R_e} + 2.5 \log{q} + 38.568,
\end{equation}
where $M_b$ is the absolute rest-frame magnitude in the B-band, $R_e$ is the half-light radius in kpc from our GALFIT fits, and $q$ is the axis ratio of the object determined by GALFIT. We computed the effective surface brightness of eGRAPES and photometric redshift galaxies and plot it as a function of redshift in Figure \ref{fig6}.
A strong monotonic decrease in the surface brightness ($\approx 2.5$ magnitudes) is visible between the redshift of  z=1.15 to z=0.2. The same effect is however observed in the non-emission line objects in the field, albeit the later have on average a much lower surface brightness than eGRAPES (by about $\approx 1 $ mag at all redshifts). The surface brightness evolution of eGRAPES does not therefore appear to be peculiar and the only distinction that eGRAPES have when compared to other galaxies in the field is that they have a very high surface brightness. Figure \ref{fig6} also shows the effect that cosmological dimming has on the measured effective surface brightness. The change in effective surface brightness caused by surface dimming is essentially flat, especially from z=0.5 to z=1.5, with a decrease of 0.6 magnitude from z=0.25 to z=1.5. Simulations showed that as objects were dimmed artificially, the measured radius decreased while the measured object flux also decreased, causing the surface brightness to remain mostly unaffected. These simulations were done by taking low redshift eGRAPES, scaling down the flux originating from the source (after masking out the background regions of the image), re-computing and adding poisson noise to the dimmed down image, adding back the original background, and re-running SExtractor and GALFIT to derived new sizes for the object.

%A more detailed comparison of the sizes of these two sets of objects is presented in Section \ref{lyasize}.  \\

\section{Size-color and Size-luminosity relations}
\begin{figure}
\includegraphics[width=7.0in]{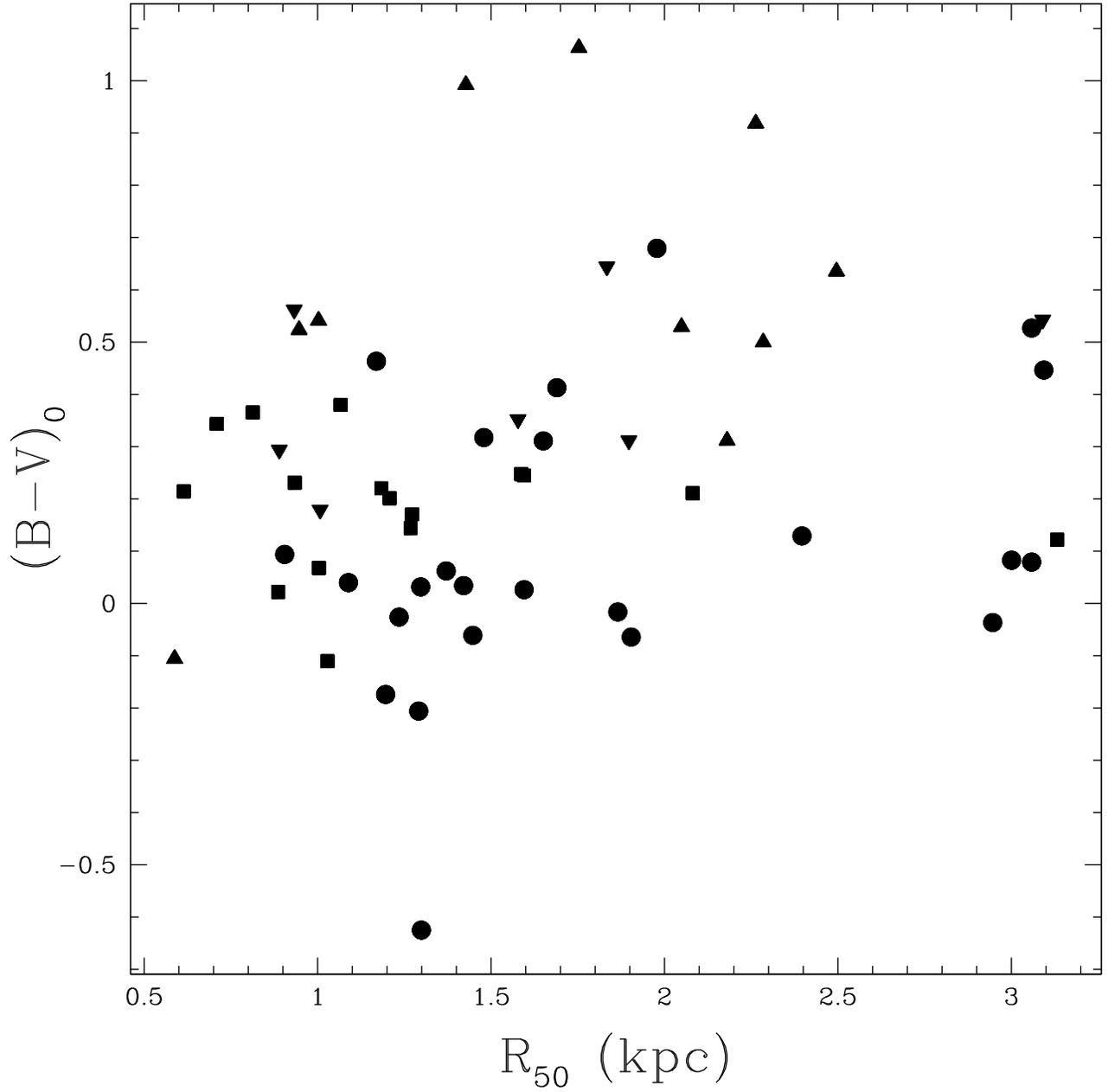}
\caption{Physical sizes and rest-frame ($\mv-\mv$) colors of eGRAPES. The filled inverted triangles, triangles, squares, and circles are objects at  $0 \leqslant z \leqslant 0.3$,  $0.3 \leqslant z \leqslant 0.55$, $0.55 \leqslant z \leqslant 0.85$, $0.85 \leqslant z \leqslant 1.5$, respectively.
\label{ReBV}}
\end{figure}

\begin{figure}
\includegraphics[width=7.0in]{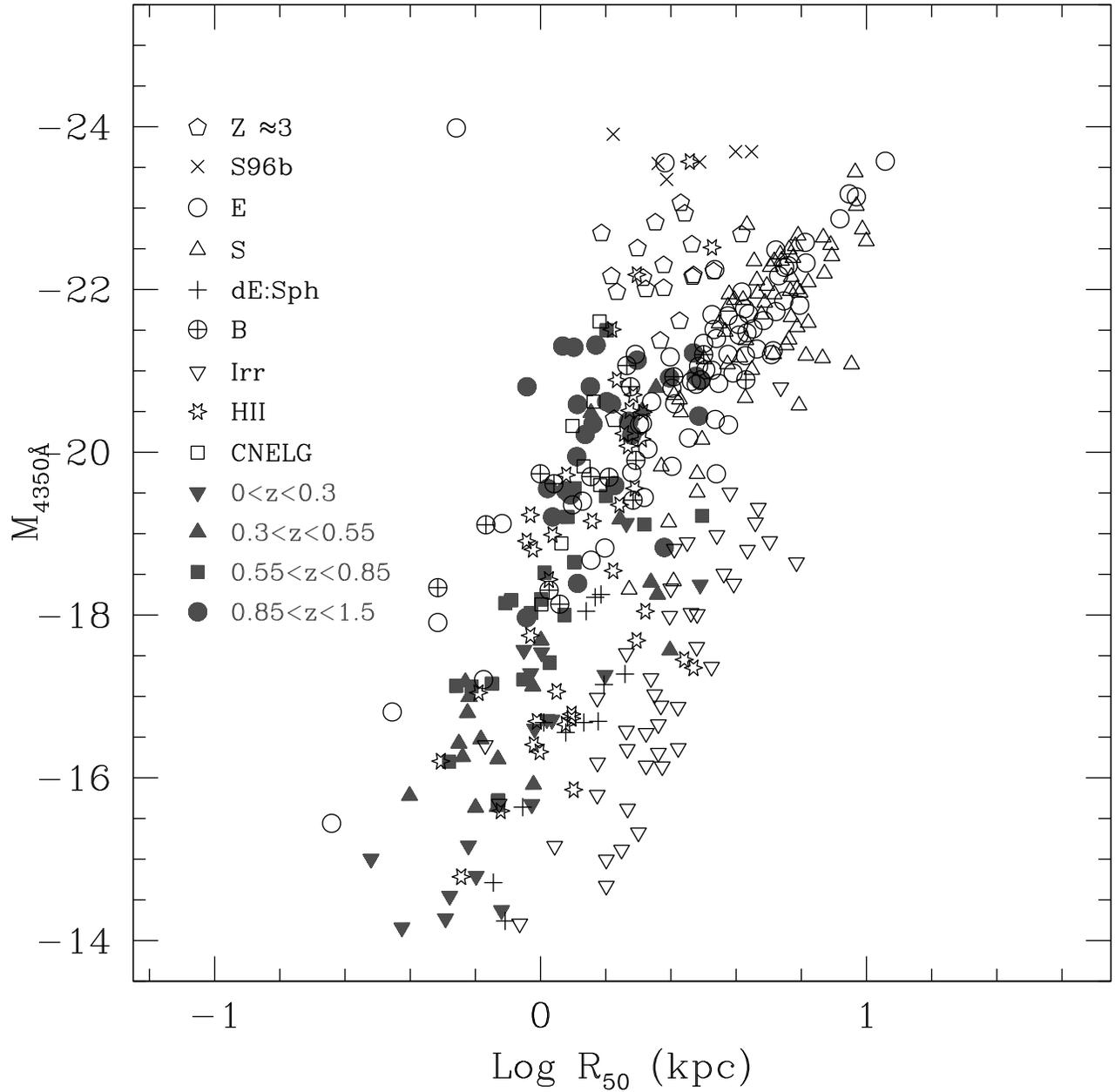}
\caption{The size and luminosity of eGRAPES (solid symbols). This Figure is similar to Figure 7 in \citet{lowenthal1996}, accounting for our choice of $h_{70}$. The open symbols are for local ellipticals, dwarfs, ellipticals/sphreroidals, spiral bulges, spirals, irregulars, HII galaxies, CNELGs, as well as the $z \approx 3$ HDF objects identified by \citet{lowenthal1996}.
\label{R50MB}}
\end{figure}

The effective surface brightness of eGRAPES increases with redshift, which could potentially be an indication of a change in the average stellar population of eGRAPES as a function of redshift. It is interesting to actually examine whether the size of eGRAPES is related to their stellar population. Figure \ref{ReBV} shows eGRAPES plotted in a rest-frame (B-V) vs $R_{50}$ plot. As this figure shows, there does not appear to be a correlation between the two. We found that eGRAPES have a median size of 1.28 kpc with sizes ranging from  $0.1 \leqslant R_{50} \leqslant 3.1$ kpc, and a median luminosity of $M_{4350\AA}=-19.8$. Figure \ref{R50MB} further shows how the size and luminosity of eGRAPES compare to local galaxies and to $z \approx 3.0$ objects from the Hubble Deep Field (UDF) \citep[][and reference therein]{lowenthal1996}. 
 %The truncation at high luminosities ($M_{4350\AA\ }\approx -21.5$) of the otherwise continuous distribution of eGRAPES is caused by the small volume probed by the HUDF observations. 
 We find that  eGRAPESs are significantly smaller and less luminous than the bulk of elliptical galaxies and slightly smaller than most of the $z \approx 3$ sources identified in the HUDF.  On the other hand, eGRAPES sizes are consistent with the observed sizes of  local irregulars, dwarfs, HII and CNELGs (Compact Narrow Emission Line Galaxies).  Yet, eGRAPESs appear to be significantly less luminous than local irregulars at any given size, or, equivalently, have significantly smaller sizes at any given luminosity than local irregulars, resulting in eGRAPESs appearing as objects with high surface brightness as shown in Figure \ref{fig6}. \\

%\section{Volume densities}
%The volume densities we derived for eGRAPES is shown in Figure \ref{vrho} using a dash line. These volume densities were computed in difference redshift bins using the $1/V_{max}$ technique and while the number of objects involved are small, we observe a slight decrease in the volume densities of eGRAPES from $z=0.5$ to $z=1.5$. Based on 11 eGRAPES , we compute a volume density of $0.016 \ {\rm h}^3_{70}\ {\rm Mpc}^{-3}$ at $z \approx 0.2$, $0.003 \ {\rm h}^3_{70}\ {\rm Mpc}^{-3}$ at $z \approx 0.7$ (14 objects), and $0.002 \ {\rm h}^3_{70}\ {\rm Mpc}^{-3}$ at $z \approx 1.2$ (14 objects). The volume density over the redshift range of $0.3 \leqslant z \leqslant 1.5$, based on a more significant number of objects (55), is $(2 \pm 0.2) \times 10^{-3} \ {\rm h}^3_{70}\ {\rm Mpc}^{-3}$.

\section{Mass Estimates}\label{SED}
We can further investigate the nature of eGRAPES by computing estimates of their mass-to-light ratios and masses. These estimates were conducted using two independent techniques. First, we used the method of \citet{bell2003}, as used by \citet{barden2005},  which relates the rest-frame color of galaxies to their SDSS r-band mass-to-light ratio and which, when using the rest-frame  (B-r) color was shown to be relatively insensitive to the detailed star formation history, metallicity, and dust content of galaxies. We computed the rest-frame (B-r) colors of eGRAPES as described above. The presence of emission line in the spectra of these objects was taken into account and results in at most a 0.05--0.10 magnitude photometric uncertainty which in turn corresponds to an error in the mass estimates of these objects of about 20\%. The mass estimates obtained in this manner are somewhat uncertain ($\approx 10-30\%$), especially for late-type objects such as eGRAPES. The results we discuss in this paper are for a population of star-forming galaxy as a whole and not for individual objects. We cannot of course follow the evolution of any single star-forming galaxy. What we can do however is to examine whether the entire population of eGRAPES has changed during the past few billions years.

The mass of eGRAPES, $\cal{M}$, allowing for the differences between the SDSS and the ACS filters \citep{fukugita1995}, can be estimated using:
\begin{equation}
\log{\cal{M}/{\rm L_r}} = -0.706 + 1.152 \times (\mb - \mv),
\end{equation}
and the stellar mass of
\begin{equation}
\log{\cal{M}} = \log{\cal{M}/{\rm L_r}} - 0.4 \times (r_s - 5 \log{D_L} - 29.67), \label{masseq}
\end{equation}
where $r_s$ is the apparent  S\'ersic magnitude at {6250\AA\ },  and $D_L$ is the luminosity distance of the object.

The second method we used  is significantly more involved and involves  the modeling of the stellar populations of each object  separately in order to determine its mass-to-light ratio from the data.  We followed a phenomenological approach describing the star formation history by a reduced set of parameters, which allowed us to scan a large range of possible star formation histories  using a two stellar component system \citep{ferreras2000}. Each of these component is a simple stellar population from the models of \citet{bruzual2003}, with a Salpeter IMF in the standard mass range ($0.1-100M_\odot$).  We assumed the same metallicity for both components, and we included the effect of dust reddening and attenuation from the prescription of \citet{charlot2000} for the younger component.

\begin{figure}
\includegraphics[width=7.0in]{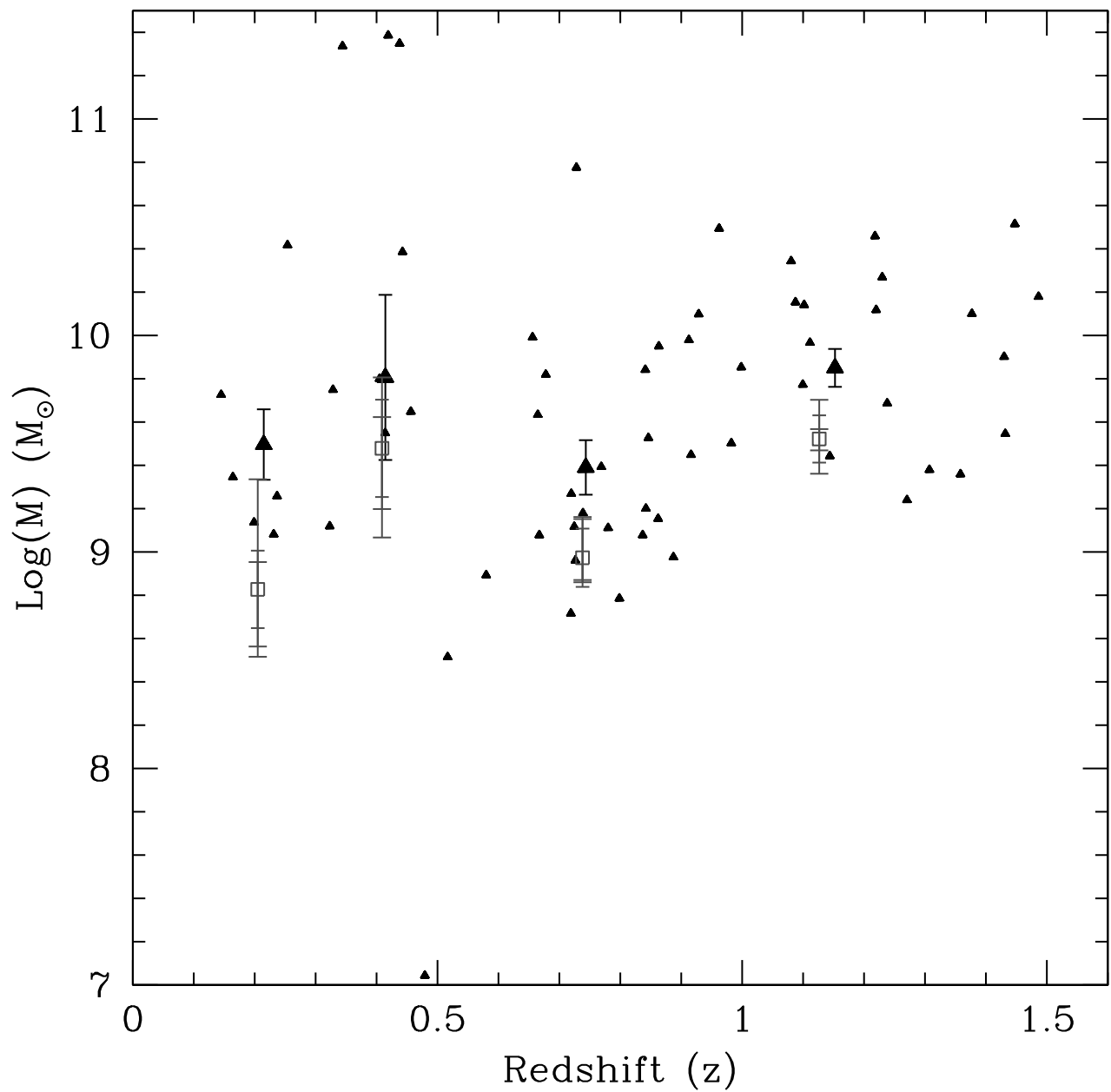}
\caption{Masses of eGRAPES  with $M_{\mb} \leqslant -17.5$. The average masses are indicated in the redshift bins of $z=0.20, 0.41, 0.73, 1.15$ using error bars. Photometric masses, as well as the three SED derived mass estimates (error bars with open symbols) described in Section \ref{SED} are plotted. The filled triangles are the individual eGRAPES photometric mass estimates.  The error bars show the 95\% confidence limit, computed using bootstrapping. \label{Mvsz}}
\end{figure}

The parameters that describe this model are the ages of both components ($t_Y$ and $t_O$); the mass ratio between the components (e.g. characterized by $f_Y$, the mass fraction in young stars); and the metallicity of both populations and the dust content -- given by $E(B-V)$.  We chose three different metallicities: $Z/Z_\odot=\{1/10,1/3,1\}$ and explored a grid of $16\times 16\times 16\times 16$ star formation histories for each metallicity, which comprises the range of parameters shown in Table \ref{table1}.
%\noindent
%Age of young component: $\log t_Y/Gyr=-3\cdots -1.$\\
%
%\noindent
%Age of old component: $t_O/Gyr = 0.5\cdots t_U(z)$, where $t_U(z)$ is the
%age of the Universe at the redshift of the galaxy. A $\Lambda$CDM concordance
%cosmology is adopted.\\
%
%\noindent
%Mass fraction in young stars: $f_Y = 0\cdots 1.$\\
%
%\noindent
%Dust (young component): $E(B-V)=0\cdots 2$~mag.
We used \mb, \mv, \mi, and \mz\ photometry from the ACS images and -- where available (86 out of 114 objects) -- NICMOS $J$ and $H$ photometry. We did not allow the eGRAPES redshifts to change and, for each galaxy, the grid of models described above was run and a  maximum likelihood method was used to determine the stellar masses. Figure \ref{Mvsz} shows the derived masses for each choice of metallicity, as well as the masses obtained using the simple photometric method described earlier. We have not included the error bars for the uncertainties expected from the color-SFH degeneracy, which amount to $\sim 0.3$~dex in $\log M/M_\odot$.

Based on the fact that the eGRAPES sample selection was limited to objects with $\mi<27$, we can estimate the lower limiting stellar mass of the eGRAPES sample by assuming a typical stellar population for these
galaxies.  Assuming an age of 0.5~Gyr, we find that the limiting eGRAPES stellar mass is $\log M/M_\odot\sim 7.6 (8.5)$ at $z=0.5 (1)$. This is well below the eGRAPES SED derived mass estimates shown in  Figure \ref{Mvsz}. Figure \ref{Mvsz} shows the individual and binned eGRAPES phometric mass estimates as well as the binned averages of the SED derived masses computed for 1/10, 1/3, and solar metallicities.The average photometric-mass of the eGRAPES population is  $ (4.5 \pm 2.1) \times 10^{9} M_\odot $. Our SED-fitting method yielded the mass estimates of
$5.1 \pm 3.5 \times 10^{9} M_\odot $, $2.4 \pm 0.5 \times 10^{9} M_\odot $, and $2.4 \pm 0.5 \times 10^{9} M_\odot $ for assuming solar, 1/3 solar, and 1/10 solar metalicities, respectively.  All of our mass estimates agree well and are essentially redshift independent. We find that eGRAPESs have masses that are similar to that of low-mass galaxies galaxies (e.g. $\leqslant 10^{10} M_\odot$) and about 10 times lower than L* galaxies today \citep{guzman1996}, independently of the metallicity we assumed for  these objects.  eGRAPES masses are similar to the mass estimates of Compact Narrow Emission Line Galaxies (CNELG) and luminous compact blue galaxies (LBG) \citep[$5 \times 10^9 M_\odot$][]{guzman1996}. These objects, which are believed to be star forming galaxies, are believed by some be the progenitors of today's more massive spiral disks galaxies \citep{phillips1997} or local dwarf elliptical galaxies \citep{guzman1996,guzman1997,guzman1998}. The relation between eGRAPESs and LCBGs is discussed further in the next section.

\begin{figure}
\includegraphics[width=7.0in]{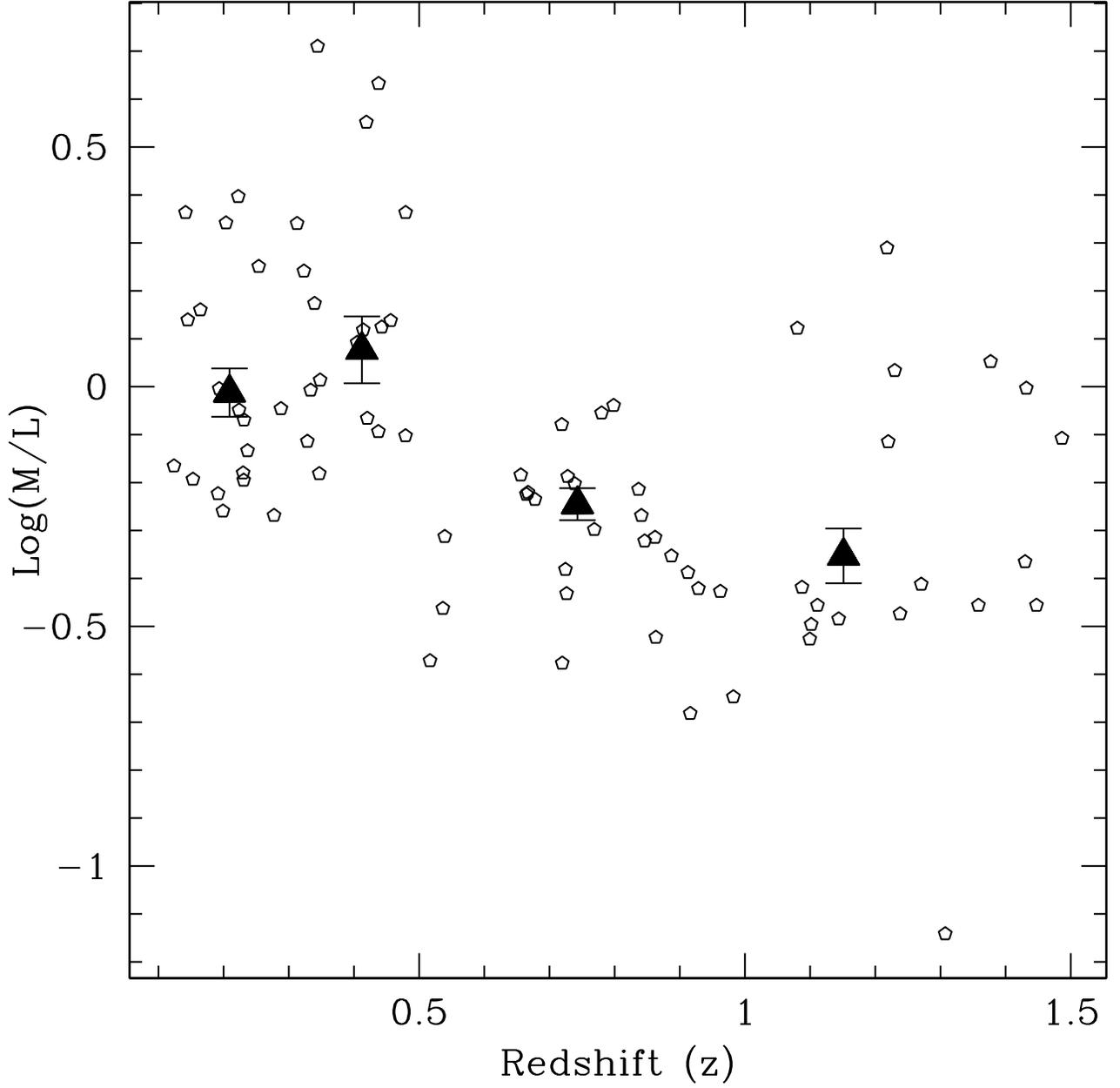}
\caption{Mass-to-light ratio of eGRAPES  with $M_{\mb} \leqslant -17.5$ (open pentagons). The average mass-to-light ratios are indicated in the redshift bins of $z=0.20, 0.41, 0.73, 1.15$ using solid triangles and error bars.  The error bars show the 95\% confidence limit, computed using bootstrapping. \label{ML}}
\end{figure}

Figure \ref{ML} further  shows the estimated mass-to-light ratio of the eGRAPES sample. The mass-to-light ratio of these objects is low ($\leqslant 1.0$), indicating that eGRAPES are objects with a young stellar population and/or that have just undergone some major starburst. Figure \ref{ML} also indicates  that the mass-to-light ratio of eGRAPES is lower at high redshifts. While we observed most of the physical attributes (such as sizes and morphology) of  eGRAPES to be uncorrelated to redshift, the increase in mass-to-light ratio at lower redshift  could be due to the fact that, on average, the observed stellar population of eGRAPES is older at lower redshifts.
This could be evidence that, on average, eGRAPES were more actively forming stars at higher redshifts, or that, alternatively, that we observe more newly born eGRAPES at high redshifts, assuming that younger galaxies have a higher fraction of mass in the form of gas and therefore a higher star formation rater per unit mass).
\\
A possible alternative would be that eGRAPES are more luminous at higher redshifts, even though they are observed to have similar sizes at all redshifts. However, while  we do observe  a  lower number of high luminosity eGRAPES  ($M_\mb \leqslant -20.0$) at low redshifts than at high redshifts, one must keep in mind that the HUDF field is small and that the volumes probed at low reshifts are small. The change of observing high luminosity eGRAPES at low redshifts  is  therefore low. For example, if we restrict ourselves to  [OII] emission line eGRAPES at a redshift $0.3 \leqslant z \leqslant 1.3$ with $M_\mb \leqslant -17.5$, we find that $60 \pm 14 \%$ eGRAPES have $M_\mb \leqslant -20.0$ at $0.8 \leqslant z \leqslant 1.3$ while  $41 \pm 15 \%$ eGRAPES have $M_\mb \leqslant -20.0$ at $0.3 \leqslant z \leqslant 0.8$. An increase in the number of high luminosity eGRAPES as a function of redshift is therefore not significantly detected in this study and the difference in the number of observed bright eGRAPES can be explained away as the result of  the relatively small volume that is probed by this study at low redshifts. It is therefore likely that the observed redshift dependence of the eGRAPES mass-to-light ratio is the result of genuinely different stellar populations.

\section{Nature of eGRAPES: Luminous compact blue galaxies?}
\begin{figure}
\includegraphics[width=7.0in]{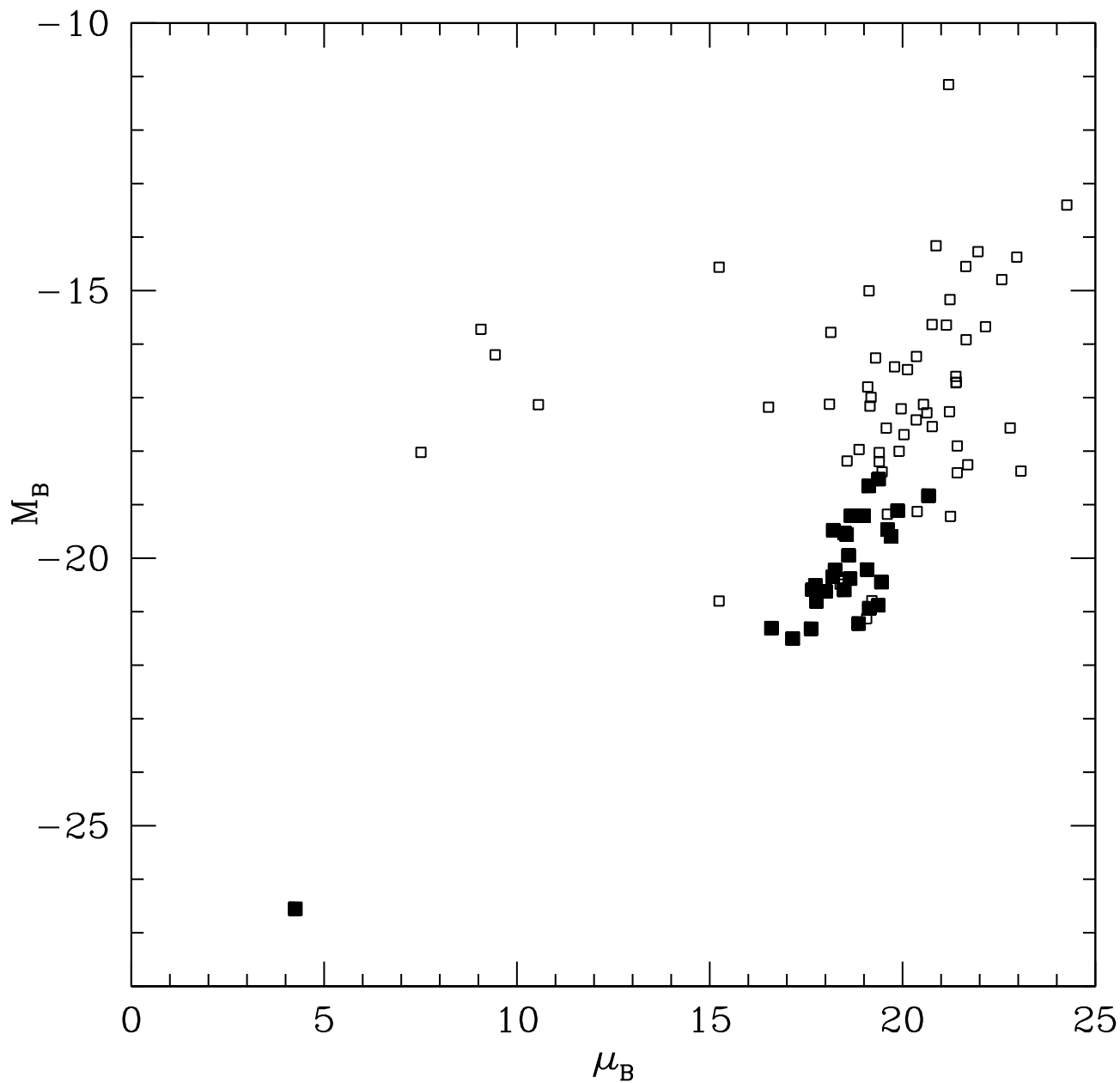}
\caption{eGRAPES in an absolute rest-frame $M_\mb\ $ magnitude versus rest-frame \mb\ band surface brightness plot. eGRAPES objects are shown using open squares. eGRAPES  luminous compact blue galaxies (LCBGs) candidates, as selected following the criteria from \citet{werk2004}, are shown using solid squares.  The very bright objects at $M_B < -25$ is a quasar, while the four objects with $\mu_B \leqslant 12$ are relatively faint, unobscured (Type 1) AGNs.\label{surfM}}
\end{figure}

We found eGRAPES to be emission line galaxies that are intrinsically very blue, compact galaxies of only a few kpc in size, and with a high surface brightness. These objects are reminiscent of a class of objects called luminous compact blue galaxies \citep[LCBG,][]{garland2004,garland2005}. These were initially identified as very blue, unresolved stellar sources in ground based QSO surveys \citep{koo1994}. LCBGs have since been shown to be a somewhat  heterogeneous group of objects composed of small star forming galaxies undergoing vigorous star formation. The exact definition of LCBGs is somewhat loosely defined and currently being refined by \citet{jangreen2005}. They include compact narrow emission line galaxies (CNELGs) \citep{koo1994,koo1995,guzman1996,phillips1997,guzman1998} and blue nucleated galaxies at larger redshifts \citep{schade1995,schade1996}. Based on the work from \citet{guzman2003,werk2004,jangreen2005}, we can identify LCBG candidates amongst the eGRAPES sample by selecting eGRAPES that have a high surface brightness (${\rm SB_e} \leqslant 21.0 {\ \rm mag\ acrsec^{-2}}$), are  blue (${\rm B-V} < 0.6$), and have a high  luminosity (${\rm M_B} \leqslant -18.5$). While eGRAPES were not initially selected based on these size, surface brightness, and luminosity cuts, Figure \ref{surfM} shows that LCBGs appear to be a natural sub-group of the eGRAPES sample. We find that approximately  60\% of  eGRAPES satisfy the \citet{werk2004} LCBG selection criteria. In Figure \ref{surfM}, the very bright object at $M_B < -25$ is likely to be a quasar and was detected in the X-ray by \cite{koekemoer2004}, while the spectra of the four objects with $\mu_B \leqslant 12$ show these objects to be  a faint QSO and three unobscured (Type 1) AGNs whose GRAPES spectra show [OIII] emission with high equivalent width.

The average volume density of eGRAPES ($M_{\mb} \leqslant -17.5$)  is $(1.8 \pm 0.3) \times
10^{-3} \ {\rm h}^3_{70}\ {\rm Mpc}^{-3}$ between $0.3 < z \leqslant
1.5$. The volume densities of the eGRAPES LCBG candidates over the same redshift range is $(1.6 \pm 0.2) \times 10^{-3} \ {\rm h}^3_{70}\ {\rm Mpc}^{-3}$ while they 
are $(2.0 \pm 0.7) \times 10^{-3}\ {\rm h}^3_{70}\ {\rm Mpc}^{-3}$ and $(2.5 \pm 0.6) \times 10^{-2}\ {\rm h}^3_{70}\ {\rm Mpc}^{-3}$ for objects at $0.4 \leqslant z \leqslant 0.7$ and $0.7 \leqslant z \leqslant 1.0$ respectively. \citet{phillips1997} estimated the LCBG volume densities to be $2.2  \times 10^{-3}\ {\rm h}^3_{75}\ {\rm Mpc}^{-3}$ and $8.8  \times 10^{-3}\ {\rm h}^3_{75}\ {\rm Mpc}^{-3}$ at $0.4 < z < 0.7$ and $0.7 < z < 1.0$, respectively,  an increase by a factor of four from low to high redshifts. If we restrict  the sample of eGRAPES LCBG candidates to objects with $M_B \leqslant -18.5$ the computed densities become $1.5  \times 10^{-3}\ {\rm h}^3_{70}\ {\rm Mpc}^{-3}$ and $1.8  \times 10^{-3}\ {\rm h}^3_{70}\ {\rm Mpc}^{-3}$ for $0.4 < z < 0.7$ and $0.7 < z < 1.0$, respectively.  More recently, \citet{werk2004} found the density of local ($z < 0.045$) LCBGs to be $5.4 \times 10^{-4}\ {\rm h}^3_{70}\ {\rm Mpc}^{-3}$.
We do not however detect a large increase in LCBG volume density as a function of redshift but  the HUDF field is however small and the numbers of eGRAPES LCBG candidates in each bins are small (6 and 12).

%The overall picture coming out this study is that eGRAPES are physically small, blue galaxies undergoing  vigorous star formation. As a whole, and while there is a large amount of individual variation in this, these objects appear to be remarkably identical at all redshifts from $0 < z \leqslant 1.5$. The  redshift identified redshift dependences that were identified were that eGRAPES appear to have  a strongly increasing $\cal{M}/{\rm L}$ ratio, that the volume number density of these objects decreases, and that the fraction of objects classified as merges increases between $0 < z \leqslant 1.5$.

%[COMMENTS ON QSOS AND AGNS]
%[COMMENT ON MASSES OF LCBGS COMPARED TO EGRAPES?]
%[EFFECT OF EMISSION LINES ON PHOTOMETRY]
%27TH MAG: FLUX? COMPARE TO LINE FLUX. USE GRAPES SPECTRA OF NON EMISSION LINE TO ESTIMATE? 
%[DO WE ONLY SEE BRIGHT LINE FLUXES AT HIGHER Z? PLOT FLUX VS Z TO ESTIMATE]
%While Section \ref{} showed that the mass-redshift distribution of the GRAPES emission line objects is flat and that on average these objects have a mass of $\approx $

\section{Conclusion}
The GRAPES survey has allowed us to select emission line galaxies,
eGRAPES, without having to first pre-select objects based on their apparent
size, luminosity, or surface brightness. We found that our spectroscopic
selection method allowed us to efficiently select a significant population of
star-forming, late type galaxies over a wide range of redshifts. We observe these
objects to be very blue (Rest-frame $(B-V) \lesssim 0.55$), to have a high surface brightness ($\approx 1$ magnitude brighter than non emission line objects in the field), small
physical sizes ($\approx 1-2\ $kpc), and relatively small masses ($\approx 5 \times 10^9 M_\odot$). We did not find any strong 
 correlation between the eGRAPES intrinsic color  and
sizes. We did observed the surface brightness of eGRAPES to increase significantly
as a function of redshift  ($\approx 2$ magnitudes between z=0.2 and z=1.15), but found no evidence that
the size of eGRAPES change with redshifts. The
mass-to-light ratio of these objects decreases as a function of redshift 
by an amount that is consistent with the observed increase in surface
brightness ($\approx 2.5$ from z=0.2 to z=1.15). This is evidence that, on average, eGRAPES were more actively forming stars at higher redshifts, or that, alternatively, that we observe more newly born eGRAPES at high redshifts.
This could be evidence that most eGRAPES have formed at higher 
redshifts. We observed that eGRAPES, while sharing many
characteristics of LCBGs and CNELGs, spanned a much wider range of
luminosities (reaching down to much lower luminosities). While the number of eGRAPES LCBG is small, we did not find a strong
redshift dependence of the volume density of these objects.

\acknowledgments
This work was supported by grant GO-09793.01-A from the Space Telescope Science Institute, which is operated by AURA under NASA contract NAS5-26555. 

\begin{deluxetable}{lr}
\tablecaption{Summary of the parameters used to model the stellar populations of each eGRAPES separately, as described in Section \ref{SED}. \label{table1}}
%%\tablehead{\colhead{Parameter}, \colhead{Range of values}}
\startdata
Age of young component, $\log(t_Y)$ (Gyr) & -3.0...-1.0 \\
Age of old component, $t_O$ (Gyr) & 0.5...$t_u(z )$\tablenotemark{1}  \\
Mass fraction in young stars, $f_Y$ & 0.0...1.0 \\
Dust (young component), E(B-V) (mag) &  0.0...2.0\\
\enddata
\tablenotetext{1}{$t_u(z)$ is the age of the Universe at the 
redshift of the galaxy. A $\Lambda$CDM concordance cosmology is adopted. }
\end{deluxetable}

%\begin{figure}
%\includegraphics[width=7.0in]{AR50}
%\caption{Asymmetry A as a function of object half-light radius in simulations. Using 166 bright UDF objects which were scaled down by factors of 2,3,4,5, and 6 before re-measuring their CAS values. {\em The A parameter remains robust as long as objects have a half-light radius that is larger than about 5 UDF pixels.} \label{AR50}}
%\end{figure}

%\begin{figure}
%\includegraphics[width=7.0in]{CR50}
%\caption{Concentration C as a function of object half-light radius in simulations. Using 166 bright UDF objects which were scaled down by factors of 2,3,4,5, and 6 before re-measuring their CAS values. {\em C only becomes significantly biased for objects smaller than 5 UDF pixels.} \label{CR50}}
%\end{figure}

%\begin{figure}
%\includegraphics[width=7.0in]{galfit}
%\caption{Redshift evolution of the volume number density of merging, extended, and concentrated eGRAPES object ($M_B \leqslant -17.5$) as a function of redshift. Membership to these three morphological classes were determined using the Galfit sersic index n as described in Section \ref{morph}  \label{galfit}}
%\end{figure}

%\begin{figure}
%\includegraphics[width=7.0in]{Vrho}
%\caption{Computed volume densities of eGRAPES (dash line) and Luminous Compact Blue Galaxies (LCBGs, solid line) in the HUDF as a function of redshift. 
%\label{vrho}}
%\end{figure}

%\begin{figure}
%\includegraphics[width=7.0in]{mum}
%\caption{Surface mass density in emission line objects as a function of redshift. The average surface mass densities are indicated in the redshift bins of $z=0.20, 0.41, 0.73, 1.15$ using error bars. \label{fig7}}
%\end{figure}

%\begin{figure}
%\includegraphics[width=7.0in]{lalphaandvdropsizes.pdf}
%\caption{z band (1500\AA rest-frame) size distribution of V drop LBG candidates and L$\alpha$ emission sources.  The two population appear to have similar $R_{50}$ sizes. The average and standard deviation of the sizes of the two populations of $z\approx 4.0$ objects are 4.71 (2.49) and 4.74 (1.73) for the L$\alpha$ emission sources and the LBG respectively.,\label{fig8}}
%\end{figure}

\end{document}